\documentclass[sn-mathphys]{sn-jnl}

\jyear{2021}%

\usepackage{graphicx}

\newcommand{\CCov}{\mathcal{C}}

\newcommand{\bbR}{\mathbb{R}}

\newcommand{\sH}{\mathcal{H}}

\newcommand{\sM}{\mathcal{M}}
\newcommand{\sR}{\mathcal{R}}
\newcommand{\sS}{\mathcal{S}}
\newcommand{\sT}{\mathcal{T}}

\newcommand{\sP}{\mathcal{P}}
\newcommand{\sU}{\mathcal{U}}

\newcommand{\vb}{\mathbf{b}}

\newcommand{\vf}{\mathbf{f}}

\newcommand{\vu}{\mathbf{u}}

\newcommand{\vx}{\mathbf{x}}
\newcommand{\vy}{\mathbf{y}}

\newcommand{\vepsilon}{\boldsymbol{\epsilon}}
\newcommand{\vzero}{\mathbf{0}}

\newcommand{\mA}{\mathbf{A}}
\newcommand{\mB}{\mathbf{B}}

\newcommand{\mD}{\mathbf{D}}
\newcommand{\mE}{\mathbf{E}}
\newcommand{\mF}{\mathbf{F}}
\newcommand{\mG}{\mathbf{G}}

\newcommand{\mI}{\mathbf{I}}

\newcommand{\mK}{\mathbf{K}}

\newcommand{\mQ}{\mathbf{Q}}
\newcommand{\mR}{\mathbf{R}}
\newcommand{\mS}{\mathbf{S}}
\newcommand{\mT}{\mathbf{T}}
\newcommand{\mU}{\mathbf{U}}
\newcommand{\mV}{\mathbf{V}}
\newcommand{\mW}{\mathbf{W}}

\newcommand{\mZ}{\mathbf{Z}}
\newcommand{\mGamma}{\boldsymbol{\Gamma}}
\newcommand{\mOmega}{\boldsymbol{\Omega}}

\newcommand{\mSigma}{\boldsymbol{\Sigma}}
\newcommand{\mPsi}{\boldsymbol{\Psi}}

\newcommand{\bcK}{\boldsymbol{\mathcal{K}}}

\newcommand{\inner}[2]{\langle #1, #2\rangle_\sH}
\newcommand{\Expect}{\mathbb{E}}
\newcommand{\intd}{\,\mathrm{d}}
\newcommand{\diag}{\mathrm{diag}}
\renewcommand{\skew}{\mathrm{Skew}}
\newcommand{\sym}{\mathrm{Sym}}
\newcommand{\grad}{\mathrm{grad}\,}
\newcommand{\tr}{\mathrm{tr}}
\newcommand{\Var}{\mathrm{Var}}

\def\trans{^\mathsf{T}}

\usepackage{amsmath}

\DeclareMathOperator*{\argmin}{arg\,min}

\usepackage{accents}

\usepackage{amsmath,amsfonts,amssymb}
\usepackage{url} 
\usepackage{ulem}
\usepackage{color}
\usepackage{caption}
\usepackage{subcaption}
\usepackage{bm}
\usepackage{multirow}

\usepackage{mathrsfs}
\usepackage{amsbsy}
\theoremstyle{thmstyleone}%
\theoremstyle{thmstyletwo}%

\theoremstyle{thmstylethree}%

\usepackage[algo2e,ruled,vlined,linesnumbered]{algorithm2e}
\usepackage[symbol]{footmisc}

\allowdisplaybreaks

\begin{document}
	
	\title[Spline Estimation of FPCs via CGPM]{Spline Estimation of Functional Principal Components via Manifold Conjugate Gradient Algorithm}
	
	\author[1]{\fnm{Shiyuan} \sur{He}
	}\email{heshiyuan@ruc.edu.cn}
	\equalcont{These authors contributed equally to this work.}
	
	\author[1,2]{\fnm{Hanxuan} \sur{Ye}}\email{hanxuan@stat.tamu.edu}
	\equalcont{These authors contributed equally to this work.}

	\author*[1]{\fnm{Kejun} \sur{He} \email{kejunhe@ruc.edu.cn}}
	
	\affil[1]{\orgdiv{The Center for Applied Statistics, Institute of Statistics and Big Data}, \orgname{Renmin University of China}, \orgaddress{\street{No. 59 Zhongguancun Street}, \city{Beijing}, \postcode{100872}, 
			\country{China}}}
	
	\affil[2]{\orgdiv{Department of Statistics}, \orgname{Texas A\&M University}, \orgaddress{\street{155 Ireland Street}, \city{College Station}, \postcode{77843}, \state{TX}, \country{USA}}}


	
	
	\abstract{Functional principal component analysis has become the most important dimension reduction technique in functional data analysis. Based on B-spline approximation, functional principal components (FPCs) can be efficiently estimated by the expectation-maximization~(EM) and the geometric restricted maximum likelihood~(REML) algorithms under the strong assumption of Gaussianity on the principal component scores and observational errors. 
	When computing the solution, the EM algorithm does not exploit the underlying geometric manifold structure, while the performance of REML is known to be unstable. In this article, we propose a conjugate gradient algorithm over the product manifold to estimate FPCs. This algorithm exploits the manifold geometry structure of the overall parameter space, thus improving its search efficiency and estimation accuracy. 
	In addition, a distribution-free interpretation of the loss function is provided from the viewpoint of matrix Bregman divergence, which explains why the proposed method works well under general distribution settings. 
	We also show that a roughness penalization can be easily incorporated into our algorithm with a potentially better fit. The appealing numerical performance of the proposed method is demonstrated by simulation studies and the analysis of a Type Ia supernova light curve dataset.}




\maketitle

\section{Introduction}\label{intro}
The statistical analysis of data in the form of curves, commonly referred to as functional data analysis \citep[FDA,][]{ramsay02}, has gained increasing research interest due to the improved capability of recording a vast amount of data in many scientific fields. 
Among the developed methodology, functional principal component analysis (FPCA) is a key technique and fundamental tool for analyzing functional data. 
It provides an informative way to explore the infinite-dimensional functional data through its variance-covariance structure and discovers features to characterize the curves \citep{ramsay05}. 
Based on the computed functional principal components (FPCs), various multivariate data analysis methods can then be applied on the principal component scores, such as regression \citep{ramsay91,yao05}, clustering \citep{james03,chiou07}, and classification \citep{james2001functional}.

Consider a random function $x(u)$ with a time index $u\in \sU$, where $\sU$ is a compact set in $\mathbb{R}$. Without loss of generality, we assume $\sU=[0,1]$. The function $x(\cdot)$ belongs to a Hilbert space $\sH \subset \mathcal{L}_2(\sU)$ equipped with the inner product  $\inner{x}{\widetilde{x}} = \int_{u\in\sU} x(u) \widetilde{x}(u)\intd u$ for $x,\widetilde{x}\in\sH$. Suppose the random element $x$ has zero mean and  covariance function $\mathcal{K}(u,v) = \Expect \{x(u) x(v) \}$. 
Under mild conditions, the Mercer's lemma states there exists an orthonormal sequence $\{\psi_{0r}(u)\}_r$  of eigenfunctions in $\sH$, and a decreasing non-negative sequence $\{\lambda_{0r}\}_{r}$ of eigenvalues, such that the covariance function can be expanded  as
\begin{equation} \label{eqn:truecov:decompose}
	\mathcal{K}(u, v) = \sum_{r=1}^\infty \lambda_{0r}\psi_{0r}(u) \psi_{0r}(v)\, .
\end{equation}
The non-negative sequence $\{\lambda_{0r}\}$ is assumed to satisfy the trace-class condition $\sum_{r=1}^\infty \lambda_{0r} <\infty$. 
Using the Karhunen-Lo{\`e}ve expansion \citep{karhunen1946spektraltheorie,loeve1946fonctions}, the random functions $x \in \sH$ can be represented as $x(u) = \sum_{r=1}^\infty \lambda_{0r}^{1/2} \theta_r \psi_{0r}(u)$, where $\theta_r$'s are uncorrelated random variables with zero mean and unit variance. Though functional data reside in infinite dimensional space, the leading $R$ eigenfunctions usually has high quality approximation power such that $\mathcal{K}(u, v) \approx \sum_{r=1}^R \lambda_{0r}\psi_{0r}(u) \psi_{0r}(v)$ and $x(u) \approx \sum_{r=1}^R \lambda_{0r}^{1/2} \theta_r \psi_{0r}(u)$. 
It is equivalent to assuming the sequence of eigenvalues $\{\lambda_{0r}\}_r$ decays fast. 
The eigenfunctions $\psi_{01},\cdots, \psi_{0R}$ are called FPCs, and the term $R$ is usually chosen such that the leading components explain a desired proportion of variability in the dataset.

In practice, we have $N$ realized functions $x_1, \cdots, x_N$ of $x$. Each function $x_n$ is discretely observed at $M_{n}$ time points $u_{n1},\cdots, u_{nM_n}\in\sU$. The observed function value $y_{nj}$ at $u_{nj}$ is corrupted by noise as 
\begin{equation} \label{eqn:signalplusnoise}
	y_{nj} = x_{n}(u_{nj} )+ \epsilon_{nj}, ~ n=1,\cdots, N, ~ j = 1,2,\cdots, M_{n},
\end{equation}
where the noises $\epsilon_{nj}$'s are independent with zero mean and finite variance $\sigma_e^2 = \Expect \epsilon_{nj}^2$. 
The goal of FPCA is to recover $\psi_{01},\cdots, \psi_{0R}$ from these $N$ sparse samples $\{(u_{nj}, y_{nj})\}_{n,j}$.

In the past decades, significant interests have been paid to the methodological and theoretical properties of FPCA. 
Some early works \cite{rice1991estimating,silverman96} constructed the estimation based on densely sampled curves. 
For sparsely observed curves, \cite{yaopca05} estimated the FPCs through learning the mean curve and the covariance surface via weighted local regression, but its convergence result was suboptimal. 
For the local smoothing method, the optimal rate was later established by \cite{hall2006properties}. More extensive theoretical investigation of the local smoothing methods has been carried out by \cite{li2010uniform} and \cite{zhang2016sparse}. 
By assuming that the sample path belongs to a reproducing kernel Hilbert space (RKHS), \cite{cai2010nonparametric} proposed an estimation based on RKHS kernels and proved its consistency. 


\begin{figure}[t]
	\centering
	\includegraphics[width = 0.8\textwidth]{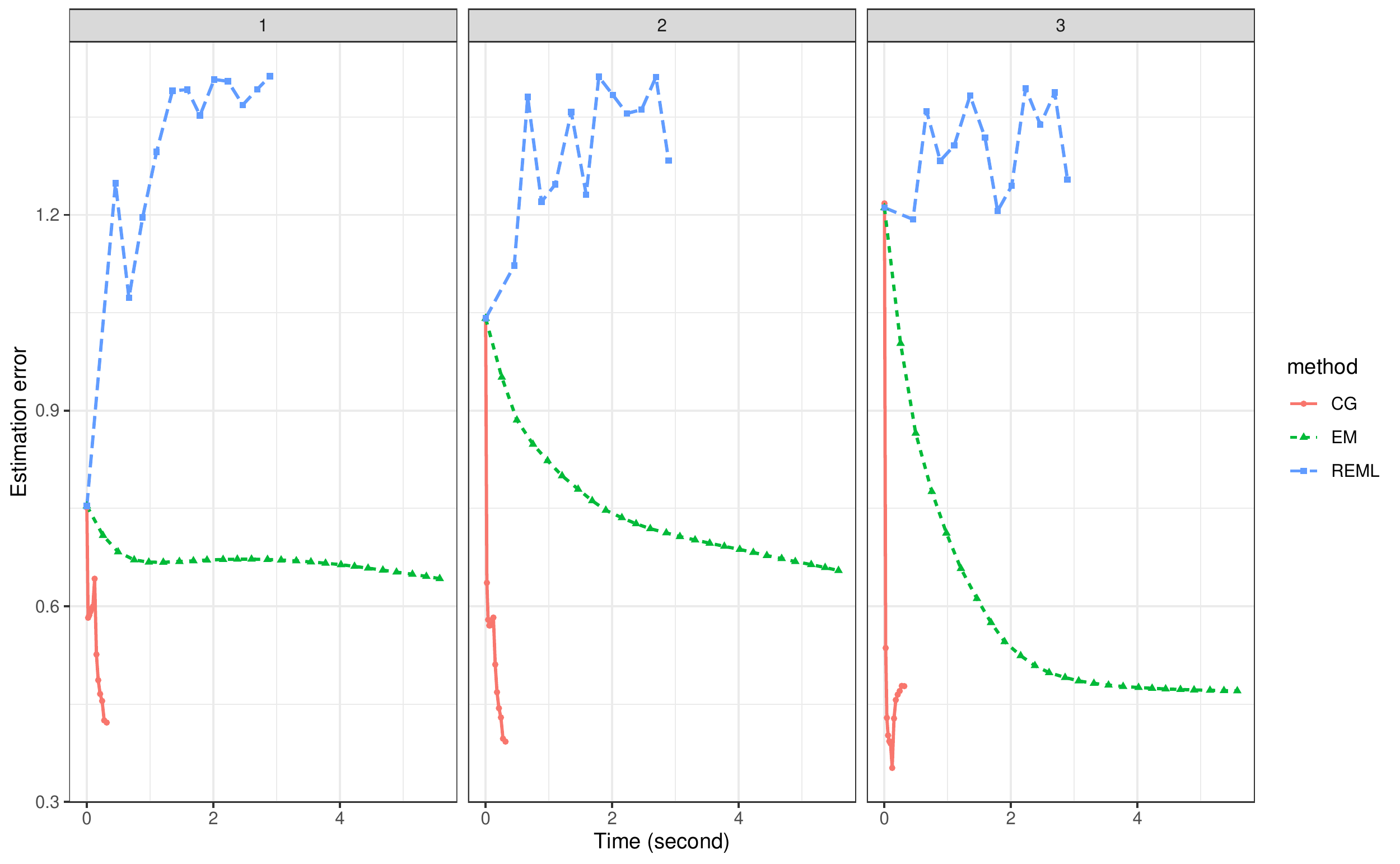}
	\caption{Plot of estimation error versus computational time for various methods in one replicate under the \textit{pracSin} simulation setting (Section \ref{sec:numerics:simulation}) with sample size $N=100$. The three panels from left to right correspond to the leading three  eigenfunctions. The $L_2$ distance between the estimated and the true eigenfunctions are shown in the vertical axis.}
	\label{fig:convergence}
\end{figure}

The above non-parametric approaches \citep{yaopca05,cai2010nonparametric} are generally computational intensive. This is because either the local regression has to be computed at a dense grid of time points \citep{yaopca05} or the number of parameters grow with the sample size \citep{cai2010nonparametric}. 
Based on spline expansion, one can establish optimization problems with a smaller number of unknown parameters \citep{james2000principal,peng2009geometric}. 
When the scores and the noises follow Gaussian distribution, the vector $\vy_n = (y_{n1}, y_{n2},\ldots, y_{nM_n})\trans$ is also multivariate Gaussian distributed.  
The estimates of the principal component functions can be obtained by maximizing the log-likelihood with spline coefficients as parameters. To solve the optimization problem, \cite{james2000principal} augmented the parameter space with the unobserved scores and used the expectation-maximization (EM) algorithm. In this optimization, the parameter space is inherently endowed with a geometric structure due to the orthonormality of the eigenfunctions and the positive semi-definiteness of the covariance function. However, the EM algorithm does not fully exploit the geometric structure during its update, except for an ad hoc post-processing step of orthogonalization at the end of each iteration to respect this structure. 
The work of \cite{peng2009geometric} explored another approach based on a geometric restricted maximum likelihood (REML). Though REML considers the manifold structure and is shown to improve accuracy over EM in \cite{peng2009geometric}, REML is found not stable and fails to converge in some replicates. 
This issue becomes worse when its initial value is not close enough to the optimal one. 
{
Figure~\ref{fig:convergence} presents the result of one replicate in the \textit{pracSin} simulation setting with sample size $N=100$. The \textit{pracSin} setting is similar the \textit{practical} setup in~\cite{peng2009geometric} such that there are $5$ eigenfunctions generated from a random combination of $10$ sinusoid basis functions (see Section 5.1 for more details).}
The figure depicts the estimation error versus computational time for EM and REML (along with our proposed method, named CG). The three panels from left to right correspond to the leading three eigenfunctions. The $L_2$ distance between the estimated and true eigenfunctions are shown in the vertical axis. In Figure~\ref{fig:convergence}, the unconvergence behavior of REML (the blue dashed curve) is evident, while EM (the green dashed curve) appears not to search very efficiently.


The purpose of this work is threefold. First, though both EM and REML are derived from the negative Gaussian log-likelihood, the estimates are consistent under broader distribution settings. In \cite{paul2009consistency}, the authors remarked that Gaussianity is more of a working assumption and can be relaxed by assuming an appropriate tail behavior of the observations. 
In this work, we show the corresponding loss function can be explained from the perspective of matrix Bregman divergence \citep{pitrik2015joint,dhillon2008matrix}, which is free of distribution assumption.  
This distribution-free motivation helps to explain why solving the loss function of EM and REML sometimes works well under more general distributions.

Second, the derived objective function with respect to spline coefficients is constrained on a rank-$R$ matrix.
Though the geometric structure of the parameter space is exploited in REML, its numerical results turn out to be unstable and heavily rely on the choice of initial value \citep{peng2009geometric}. 
Through looking insight into the manifold geometry \citep{absil2009optimization,hu2020brief}, we propose a conjugate gradient algorithm for FPC estimation according to the recent advances in matrix manifold optimization.
In particular, we factorize the fixed rank-$R$ symmetric matrix as a product of two matrices, such that one belongs to the Stiefel manifold \citep{edelman1998geometry} and the other is within the positive definite cone  \citep{sra2015conic}. In contrast to \cite{peng2009geometric}, we view the two matrices together as an element of the corresponding product manifold, and the conjugate gradient algorithm is deployed over this product structure by taking into account its intrinsic geometry. 
Figure~\ref{fig:convergence} also plots the convergence trace of our proposed method (CG) in solid red curve. The improvement over REML and EM is substantial. 

Our third task is to compare two approaches of tuning empirically. One is using regression splines where the tuning parameter is the number of knots; the other is using penalized splines with a roughness penalty parameter to be tuned. 
Both REML and EM are based on regression splines with equally spaced knots within the domain, and the number of knots needs to be tuned. 
One drawback of this approach is the discrete tuning in the sense that the number of knots is selected from a set of discrete integers, which may result in unstable estimation.
We suggest using penalized splines \citep{o1986statistical, o1988fast,huang2021asymptotic} by fixing the number of knots at a relatively large value and incorporating the roughness penalization to achieve continuous tuning.
Our optimization algorithm allows us to directly integrate the roughness penalty, and we find that the penalized spline estimator potentially provides more accuracy in estimation and prediction compared with the one using regression splines.


The rest of the article is organized as follows. Our spline estimator is motivated from the matrix Bregman divergence, and the corresponding optimization problem is formulated in Section~\ref{sec:meth}. 
We review basic concepts of product manifold geometry in Section~\ref{sec:prodMan}. Based on these concepts, a matrix manifold conjugate gradient algorithm is proposed in Section~\ref{sec:algorithm}. 
In Section~\ref{sec:numerics}, our method is compared with some existing alternatives using both simulated and real datasets. 
{
This paper is summarized in Section \ref{sec:discuss} with a brief concluding remark.}

\section{Methods} \label{sec:meth}

This section formulates the optimization problem to estimate the leading $R$ FPCs from sparsely sampled  data. 
Our method is built upon a rank-$R$ covariance function $\CCov$, defined in a tensor product spline space, to approximate the true covariance function $\mathcal{K}$ in \eqref{eqn:truecov:decompose}. 
In particular, over the domain $\sU$, we let $\vb(\cdot)\in\bbR^K$ be an orthonormalized spline basis 
with degrees of freedom $K$. 
The orthonormality means that  $\int \vb(u)\vb\trans(u)\intd u = \mI\in\bbR^{K\times K}$.
{
This orthonormality requirement facilitates the interpretation and computation of the functional principal component by transforming the $L_2$ norms (or inner products) of eigenfunctions to the Euclidean norms (or inner products respectively) of their  basis coefficient vectors.
Meanwhile, the choice of spline basis order is closely related to the smoothness of the underlying eigenfunctions. 
We suggest using the cubic splines ($4$th order) as in \cite{james2000principal,huang2021asymptotic} when the prior information about the  eigenfunction smoothness is unknown.} 
Given the vector of basis functions, we model the covariance function as $\CCov(u,v)=\vb\trans(u)\mD\vb(v)$ using a matrix $\mD\in\bbR^{K\times K}$ which is positive semi-definite with rank $R$.  
Suppose its eigen-decomposition is $\mD = \mU\mW\mU\trans$, where $\mU\in\bbR^{K\times R}$ has orthonormal columns and $\mW=\diag(\lambda_{1},\cdots, \lambda_{R})\in\bbR^{R\times R}$ is a diagonal matrix. 
Let $\vu_r$ be the $r$-th column of $\mU$ and denote $\psi_r(u) = \vu_r\trans\vb(u)$. 
It can be shown straightforwardly that 
\begin{equation} \label{eqn:modeleigenfun:delta}
	\int \psi_r(u)\psi_{r'}(u) \intd u = \vu_{r}\trans \Big\{\int \vb(u)\vb\trans(u)\intd u\Big\} \vu_{r'}= \vu_{r}\trans\vu_{r'} = \delta_{rr'},
\end{equation}
where $\delta_{rr'}$ is the Dirac delta, and that 
\begin{equation*} \label{eqn:modeleigenfun:eigen}
	\begin{aligned}
		\int \CCov(u,v) \psi_r(v) \intd v &= \vb(u)\trans \mU \mW \mU\trans \Big\{\int  \vb(v)\vb(v)\trans \intd v\Big\}\vu_{r} \\
		&= \vb(u)\trans \mU \mW \mU\trans \vu_{r} \\
		&= \lambda_{r} \vb(u)\trans \vu_{r} \\
		& = \lambda_{r}\psi_{r}(u).
	\end{aligned}
\end{equation*}
Thus, written in this way, the modeled covariance function $\CCov$ has $\lambda_{r}$ as its eigenvalue, and the corresponding eigenfunction is $\psi_r(u) = \vu_r\trans\vb(u)$, for $r=1,\ldots, R$. 

For the $n$-th sample $\vy_n = (y_{n1},\cdots, y_{nM_n})\trans$ with $y_{nj}$ in \eqref{eqn:signalplusnoise},
$n=1,\cdots, N$ and $j = 1,2,\cdots, M_{n}$, its covariance matrix is
\begin{equation} \label{eqn:trueCovMat}
	\bcK_n = \big[\mathcal{K}(u_{nj}, u_{nj'})\big]_{j,j'} + \sigma_e^2\mI.
\end{equation}
Due to our spline representation of the covariance function $\CCov(u,v)=\vb\trans(u)\mD\vb(v)$ with $\mD = \mU\mW\mU\trans$, $\bcK_n$ in \eqref{eqn:trueCovMat} will then be approximated by 
\begin{equation}\label{eqn:modelCovMat}
	\mSigma_n = \big[
	\CCov(u_{uj}, u_{nj'})\big]_{jj'} + \sigma_e^2\mI,
\end{equation}
with parameters $\mU$, $\mW$, and $\sigma_e^2$. 
To estimate the unknown parameters $\mU$, $\mW$ and $\sigma_e^2$, we propose a loss function based on the matrix Bregman divergence \citep{pitrik2015joint,dhillon2008matrix} to measure the discrepancy between the covariance matrices $\bcK_n$ in \eqref{eqn:trueCovMat} and $\mSigma_n$ in \eqref{eqn:modelCovMat}.  

A matrix Bregman divergence  is constructed with  a strictly convex and twice continuously differentiable function $\varphi(\cdot):\bbR\to \bbR$. 
This function can be applied to the positive definite matrix $\mSigma_n \in\bbR^{M_n \times M_n}$ via its eigendecomposition $\mSigma_n = \mF_n\mG_n\mF_n$, where the columns of $\mF_n$ contain the eigenvectors of $\mSigma_n$ and $\mG_n = (g_{n1},\cdots, g_{nM_n})$ is the diagonal matrix of its eigenvalues. 
We denote $\varphi(\mSigma_n) = \mF_n\varphi(\mG_n)\mF_n\trans$, where $\varphi(\mG_n) = \diag\{\varphi(g_{n1}),\cdots, \varphi(g_{nM_n})\}$ such that $	\varphi$ is applied elementwisely to the diagonal. 
The function $\varphi$ induces a matrix Bregman divergence for the pair of positive definite matrices
$\bcK_n$ and $\mSigma_n$ as 
\begin{equation} \label{eqn:define:divergence}
	\mathcal{D}_{\varphi}(\bcK_n \| \mSigma_n) = \tr\big\{\varphi(\bcK_n) - \varphi(\mSigma_n) -   \varphi'(\mSigma_n)(\bcK_n - \mSigma_n)\big\},
\end{equation}
and we call $\varphi$ the \textit{seed} function.
This divergence gauges the similarity between the true covariance matrix $\bcK_n$ and our model covariance $\mSigma_n$, and \eqref{eqn:define:divergence} is zero if and only if $\bcK_n=\mSigma_n$.

In this work,  we choose the seed function $\varphi(x) = -\log(x)$, and the divergence loss becomes
the LogDet divergence
\begin{equation} \label{eqn:define:divergence2}
	\mathcal{D}_{\varphi}(\bcK_n \| \mSigma_n) = \tr\big\{ -\log(\bcK_n) + \log(\mSigma_n) + \mSigma_n^{-1} (\bcK_n - \mSigma_n)\big\}.
\end{equation}
Note also the true covariance matrix $\bcK_n$ is not observed in practice. Fortunately, we have its one-sample estimate $\mS_n = \vy_n\vy_n\trans$, which is an unbiased estimate of $\bcK_n$
{
at the same observational points $\{u_{nj}:j=1,\cdots, M_n\}$ of $\vy_n$,} 
$n=1,\cdots, N$. 
{Notice that the functional data are assumed to be sparsely observed, and the observational time points $\{u_{nj}:j=1,\cdots, M_n\}$ are randomly drawn from the domain $\mathcal{U}$. This means, for the $n$-th sample, the random observational time points $\{u_{nj}:j=1,\cdots, M_n\}$ are unique to this particular sample. The one-sample estimate $\mS_n$  therefore provides the only available information for the corresponding true covariance matrix $\bcK_n$ in~\eqref{eqn:trueCovMat} at the observational time points of $\vy_n$.}
Furthermore, the right hand side of~\eqref{eqn:define:divergence2} indicates that the first term $\log(\bcK_n)$ is fixed with respect to the parameters. 
Therefore, discarding the first term $\log(\bcK_n)$ and replacing the last $\bcK_n$ by its one-sample estimate $\mS_n$, we get the loss function over the $N$ samples \allowdisplaybreaks
\begin{align}\label{eqn:loss:logdet}
	\mathcal{L}(\mU,\mW,\sigma_e^2)  
	&=	\frac{1}{N}\sum_{n=1}^{N}
	\big\{\log\det \mSigma_n +\langle \mSigma^{-1}_n,\, \mS_n -\mSigma_n\rangle \big\} \nonumber \\
	&= \frac{1}{N}\sum_{n=1}^{N}
	\big\{\log\det \mSigma_n +\langle \mSigma^{-1}_n,\, \mS_n \rangle \big\} + C,
\end{align}
where $C$ is a relative constant.  From the above, we can see this loss corresponds to the negative log-likelihood of $\vy_n$ when the data is assumed to follow multivariate Gaussian distribution. The same loss has been used in~\cite{james2000principal} and \cite{peng2009geometric} where Gaussian distribution was assumed. The above derivation motivates that the loss function works under more general settings without assuming Gaussianity.

Given $N$ sparse samples and the loss function~\eqref{eqn:loss:logdet}, the estimation can be obtained via solving
\begin{equation}\label{eqn:firstOptimization} 
	\begin{aligned}
		\argmin_{\mU, \mW,\sigma_e^2}\, & \mathcal{L}(\mU,\mW,\sigma_e^2), \\
		\text{subject to }\, & \mU\in \mathrm{St}(R, K),\;
		\mW = \diag(\lambda_{1},\cdots, \lambda_{R})\succ \vzero, \text{ and } \sigma_e^2 >0 ,
	\end{aligned}
\end{equation}
where $\mathrm{St}(R, K)=\{\mU\in\bbR^{K\times R}:\, \mU\trans\mU = \mI\}$ is the Stiefel manifold \citep{edelman1998geometry} and all the eigenvalues $\lambda_{r}$ are required to be non-negative. 
In \eqref{eqn:firstOptimization}, the noise variance $\sigma_e^2$ is also a parameter to be estimated. 
Denote $(\hat{\mU}, \hat{\mW})$ and $\hat{\sigma}^2_e$ as the optimal solution to problem~\eqref{eqn:firstOptimization}. 
The $r$-th eigenfunction is then estimated by $\hat{\psi}_{r}(u) = \vb\trans(u)\hat{\vu}_{r}$, where $\hat{\vu}_{r}$ is the $r$-th column of  $\hat{\mU}$. 
Requiring $\mU$ to  belong to the Stiefel manifold $\mathrm{St}(R, K)$ makes the estimates $\hat{\psi}_{r}(u)$ compatible with the eigenfunctions orthonormality property~\eqref{eqn:modeleigenfun:delta}.

\section{Geometry of the Product Manifold}\label{sec:prodMan}

Before presenting the algorithm to solve~\eqref{eqn:firstOptimization}, we first note that $\mU$ and $\mW$ belong to the Stiefel manifold $\text{St}(R, K)$ \citep{edelman1998geometry} and the cone of positive definite matrices $\sS_{++}(R)$ \citep{smith2005covariance,sra2015conic}, respectively. 
The factored matrices $(\mU,\mW)$ are identifiable up to a column/row permutation. The diagonal constraint on $\mW$  in~\eqref{eqn:firstOptimization} can be too restricted for an efficient search of $(\mU,\mW)$. 
We consider an enlarged parameter space for the possibility of a shorter path via which the optimal value is reached faster.
In particular, we let $\sM$ denote the product manifold \citep{absil2009optimization} of $\text{St}(R, K)$ and $\sS_{++}(R)$, and treat $(\mU, \mW)$ as a point on $\sM$, i.e., 
$$
x=(\mU, \mW) \in \sM= \text{St}(R, K)\times \sS_{++}(R).
$$
The above $\sM$ is a submanifold of $\bbR^{K\times R}\times \bbR^{R\times R}$ and its manifold topology is equivalent to the product topology \citep{absil2009optimization} of $\text{St}(R, K)$ and $\sS_{++}(R)$.
To improve the search efficiency of parameter $x = (\mU,\mW)$, the geometrical aspects of $\sM$ need to be taken into account, and we present the geometric structures of the product manifold in this section.

\paragraph{The tangent space.}As a product manifold, the tangent space at $x = (\mU,\mW) \in \sM$ is denoted as 
\begin{equation*}
	\sT_{x}\sM = \sT_{\mU} \mathrm{St}(R, K) \times \sT_{\mW} \sS_{++}(R),
\end{equation*} 
where $\sT_{\mU} \mathrm{St}(R, K)$ and $\sT_{\mW} \sS_{++}(R)$ are the tangent spaces of $\mathrm{St}(R, K)$ at $\mU$ and $\sS_{++}(R)$ at $\mW$, respectively. 
In other words, each tangent vector $\bar{\xi}_{x} = (\bar{\xi}_{\mU} , \bar{\xi}_{\mW})$ has two components: one is $\bar{\xi}_{\mU}$ in the tangent space of Stiefel manifold $\sT_{\mU} \mathrm{St}(R, K)$, and the other is $\bar{\xi}_{\mW}$ in the tangent space of positive definite manifold $\sT_{\mW} \sS_{++}(R)$. Both tangent spaces have explicit forms, i.e.,
\begin{align*}
\sT_{\mU} \mathrm{St}(R, K) &= \big\{\mU\mOmega+\mU^{\perp}\mK:\ \mOmega\trans=-\mOmega,\ \mK\in\bbR^{(K-R)\times R}\big\} \\
\sT_{\mW} \sS_{++}(R) & = \big\{\mK\in\bbR^{R\times R}: \mK=\mK\trans\big\}.
\end{align*}
In the above, $\mU^{\perp}\in\bbR^{K\times(K-R)}$ is any matrix whose column space is the orthogonal complement of $\text{span}(\mU)$. The tangent space $\sT_{\mW} \sS_{++}(R)$ is simply the set of symmetric matrices of size $R$.

\paragraph{The gradient of a function.} In order to facilitate the direction corresponding to the steepest descent from $x\in \sM$, we further characterize the length on the tangent vectors. 
This is done by endowing $\sT_{x} \sM$ with a metric. As for the product of manifolds $\text{St}(R, K)$ and $\sS_{++}(R)$, the endowed inner product of $\sT_{x} \sM$ is defined as 
\begin{equation}\label{eqn:prodManTanMet}
	g_{x}(\bar{\xi}_{x}, \bar{\eta}_{x}) =
	\langle \bar{\xi}_{\mU} , \bar{\eta}_{\mU}\rangle +\,
	\langle \bar{\xi}_{\mW}\mW^{-1} , \mW^{-1}\bar{\eta}_{\mW}\rangle\, ,
\end{equation}
for two tangent vectors $\bar{\xi}_{x}= (\bar{\xi}_{\mU} , \bar{\xi}_{\mW})\in \sT_{x} \sM$ and $\bar{\eta}_{x}= (\bar{\eta}_{\mU}$, $\bar{\eta}_{\mW})\in \sT_{x} \sM$ \citep{meyer2011geometric}.
On the right hand side of \eqref{eqn:prodManTanMet}, the first term is the matrix inner product inherited from the Euclidean space for $\mathrm{St}(R,K)$, and the second term is the intrinsic inner product for $\sS_{++}(R)$. For a generic smooth function $F$ on the Euclidean space $\bbR^{K\times R}\times\bbR^{R\times R}$, we let $\bar{f}(x)$ be the corresponding function when the domain of $F$ is restricted to $\sM$, i.e., 
$$
\bar{f}(x) = F(\mU,\mW), \mbox{ for } x = (\mU,\mW)\in \sM.
$$
Using \eqref{eqn:prodManTanMet}, the \textit{gradient} of $\bar{f}$, denoted as $\grad\bar{f}$, is the unique element in $\sT_{x}\sM$ such that 
$$
g_{x}(\grad\bar{f}, \bar{\xi}_{x})
= \Big\langle \frac{\partial F}{\partial \mU}, \bar{\xi}_U\Big\rangle + 
\Big\langle \frac{\partial F}{\partial \mW}, \bar{\xi}_W\Big\rangle 
$$
for all $\bar{\xi}_{x}\in \sT_{x} \sM$ \citep{absil2009optimization}. 

After calculating the partial derivatives ${\partial F}/{\partial \mU}$ and ${\partial F}/{\partial \mW}$ in the Euclidean space, we can evaluate the manifold gradient as $\grad\bar{f}= \Psi_{x}({\partial F}/{\partial \mU}, {\partial F}/{\partial\mW})$, where the projection $\Psi_{x}$ is defined as
\begin{equation}\label{eqn:projectTangent}
	\begin{aligned}
		&\Psi_{x}:\ \bbR^{K\times R}\times \bbR^{R\times R} \to
		\sT_{x}\sM,\, \\
		&\qquad \big(\mZ_\mU, \mZ_\mW\big) \mapsto 
		\big((\mI - \mU\mU\trans)\mZ_\mU + \mU
		\skew (\mU\trans\mZ_\mU), \mW\sym(\mZ_\mW) \mW\big) \, . 
	\end{aligned}
\end{equation}
In the above, we denote $\skew(\mA) = (\mA-\mA\trans)/2$ and $\sym(\mA) =(\mA+\mA\trans)/2$ for a generic matrix $\mA$.

\paragraph{Vector transport.} 

Unlike the canonical metric in a Euclidean space, the tangent vectors at different points of a manifold cannot be directly added since they are indeed in different tangent spaces. This phenomenon raises a challenge when we solve an optimization problem over a manifold by moving the iteration point to the next.  
To overcome this difficulty, we now investigate a notation named \textit{vector transport} \citep{edelman1998geometry} that is consistent with the metric we defined in \eqref{eqn:prodManTanMet}.
In particular, a vector transport $\sP_{x \to y}$ moves a tangent vector $\bar{\xi}_{x} \in \sT_{x}\sM$ at a point $x=(\mU, \mW)$ to the tangent space 
$\sT_{ y}\sM$ at another point $ y = (\widetilde{\mU},\widetilde{\mW})$. 
For the Stiefel manifold, the tangent vector projection (the first part of~\eqref{eqn:projectTangent}) can be employed \citep{absil2009optimization}. 
For the positive definite manifold, we use its exact \textit{parallel transport} \citep{edelman1998geometry} which moves the tangent vector along its geodesics.
Formally, the vector transport is defined as
\begin{equation*}
	\sP_{x \to y}: \sT_{x}\sM\to
	\sT_{ y}\sM,\quad 
	\bar{\xi}_{x} = (
	\bar{\xi}_{\mU}, \bar{\xi}_{\mW} )
	\mapsto \bar{\xi}_{ y} = (
	\bar{\xi}_{\widetilde{\mU}}, \bar{\xi}_{\widetilde{\mW}} ),
\end{equation*}
where the components of the mapped $\bar{\xi}_{ y}$ are computed through
\begin{align*}
	&  \bar{\xi}_{\widetilde{\mU}} = (\mI - \widetilde{\mU}\widetilde{\mU}\trans)\bar{\xi}_\mU + \widetilde{\mU}
	\skew (\widetilde{\mU}\trans \bar{\xi}_\mU), \\ 
	& \bar{\xi}_{\widetilde{\mW}} =\mW^{{1}/{2}}
	(\mW^{-{1}/{2}} \widetilde{\mW}\mW^{-{1}/{2}})^{{1}/{2}}
	\mW^{{1}/{2}}\bar{\xi}_W \mW^{{1}/{2}}
	(\mW^{-{1}/{2}} \widetilde{\mW}\mW^{-{1}/{2}})^{{1}/{2}}
	\mW^{{1}/{2}} . 
\end{align*}

\paragraph{Retraction.} The last geometric element used in manifold optimization is \textit{retraction} $\sR_{x}$. 
Retraction ensures the point $x$ remains on the manifold after moving along the search direction in the tangent space \citep{edelman1998geometry}, i.e., it is a smooth mapping from the tangent space $\sT_{x}\sM$ onto the manifold $\sM$ and defined as
\begin{equation}\label{eqn:subRectraction}
	\begin{aligned}
		& \sR_{x}: \sT_{x}\sM\to \sM,\ 
		\bar{\xi}_{x}=(\bar{\xi}_{\mU},\bar{\xi}_{\mW}) \mapsto x' = (\mU', \mW')\, \mbox{ with}
		\\
		&\qquad  \mU'= \mathrm{qr}(\mU + \bar{\xi}_\mU),  \\ 
		&\qquad  \mW' =  \mW^{1/2}\exp(\mW^{-1/2} \bar{\xi}_\mW \mW^{-1/2})
		\mW^{1/2} \,.
	\end{aligned}
\end{equation}
In \eqref{eqn:subRectraction}, $\mathrm{qr}(\cdot)$ extracts the Q component of a QR decomposition. Specifically, with QR decomposition of a matrix $\mV = \mQ\mR$, we denote $\mathrm{qr}(\mV) = \mQ$, and the diagonal elements of  $\mR$ are ensured to be positive. 
The third line of \eqref{eqn:subRectraction} is indeed the exponential mapping \citep{absil2009optimization} for the positive definite manifold $\sS_{++}(R)$, where $\exp(\cdot)$ is the matrix exponential. 

We end this subsection by remarking that the working domain $\sM = \text{St}(R, K)\times \sS_{++}(R)$ is larger than the original constraint  of~\eqref{eqn:firstOptimization}, where the matrix $\mW$ is restricted to be a diagonal matrix with positive elements. 
This positive diagonal structure will be recovered after updating $\mU$ and $\mW$ (see Algorithm~\ref{algorithm:CG} in Section \ref{sec:algorithm}). 

\section{Conjugate Gradient Algorithm over the Product Manifold} \label{sec:algorithm}

We now present how to solve~\eqref{eqn:firstOptimization} by alternatively updating $(\mU,\mW)$ and $\sigma^2_e$ in a cyclic manner. 
In other words, with fixed $\sigma_e^2$, the pair of matrices $(\mU,\mW)$ gets updated; then, with fixed $(\mU, \mW)$, the noise variance $\sigma_e^2$ gets updated. The latter step is relatively straightforward, as $\sigma_e^2$ can be updated by a classical gradient method. 
For updating $(\mU,\mW)$, we propose to use conjugate gradient algorithm \citep{nocedal2006numerical} over the product manifold $\sM$ and will present it afterward. 
Algorithm~\ref{alg:outerloop} summarizes the main steam.

\begin{algorithm2e}[!h] 
	\KwIn{Initial value for $\mU,\mW,\sigma_e^2$.}
	\KwOut{The estimtaes of $\mU,\mW,\sigma_e^2$. }
	\BlankLine
	\DontPrintSemicolon
	\Repeat{convergence.}{
		For fixed $\sigma^2_e$,  update $(\mU,\mW)\in\sM$ by
		the conjugate gradient algorithm in Algorithm~\ref{algorithm:CG}.\;
		For fixed $(\mU,\mW)$, update $\sigma^2_e$ by a classical gradient method.\;		
	}
	\caption{The Main Algorithm. \label{alg:outerloop}}
\end{algorithm2e}

The conjugate gradient algorithm to update $(\mU,\mW)$ is presented in Algorithm~\ref{algorithm:CG} aiming to minimize \eqref{eqn:firstOptimization} efficiently.
Compared with the steepest descent, the conjugate gradient algorithm accelerates the convergence rate \citep{bertsekas1999nonlinear,nocedal2006numerical} 
{
by adding an additional term to ensure conjugacy}. Meanwhile, unlike the Newton's method, the conjugate gradient algorithm also reduces the cost of numerical iterations by avoiding evaluating the inverse of the Hessian operator \citep{bertsekas1999nonlinear}.  Algorithm~\ref{algorithm:CG} works on the product manifold and relies on two major components. The first is to exploit the manifold structure to update $x$, which has been discussed in Section~\ref{sec:prodMan}.
The second is computing the partial derivatives~\eqref{eqn:partial:FU} and~\eqref{eqn:partial:FW} of $F$ in the Euclidean space. 
In particular, set $F(\mU,\mW):=\mathcal{L}(\mU,\mW,\sigma_e^2)$ for the loss function~\eqref{eqn:loss:logdet} in the Euclidean space with $\sigma_e^2$ fixed. 
By a direct application of the chain rule, the partial derivative of $F$ w.r.t. $\mU$ and $\mW$ are
\begin{align}
	\frac{\partial F(\mU, \mW)}{\partial \mU}
	&= \frac{1}{N} \sum_{n=1}^{N} 
	2\mB_n\trans\bigg\{ \frac{\partial L_n (\mSigma_n)}{\partial \mSigma_n}\bigg\}\mB_n  \mU\mW \label{eqn:partial:FU}
\end{align}
and
\begin{align}
	\frac{\partial F(\mU, \mW)}{\partial \mW}
	&= \frac{1}{N} \sum_{n=1}^{N} 
	\mU\trans\mB_n\trans\bigg\{ \frac{\partial L_n (\mSigma_n)}{\partial \mSigma_n}\bigg\}  \mB_n\mU, \label{eqn:partial:FW}
\end{align}
where 
$$\frac{\partial L_n (\mSigma_n)}{\partial \mSigma_n} = \mSigma_n^{-1}-  \mSigma_n^{-1}\mS_n \mSigma_n^{-1}.$$
To be consistent with Section~\ref{sec:prodMan}, we use $\bar{f}$ to denote $F$ when the domain is restricted to the product manifold $\sM$.

\begin{algorithm2e}[t]
	\KwIn{Initial iterate $x_0 =(\mU_0,\mW_0) \in \sM$.}
	\KwOut{Computed estimations for ${\mU}$ and ${\mW}$ when $\sigma_{e}^2$ is fixed. }
	\BlankLine
	\DontPrintSemicolon
	Set $k=0$. \;
	Compute $\frac{\partial F}{\partial \mU}$ and $\frac{\partial F}{\partial  \mW}$ at the current $x_0$. \;
	Compute the gradient $\mathrm{grad} \bar{f}(x_0) = \Psi_{x_0} (\frac{\partial F}{\partial \mU}, \frac{\partial F}{\partial  \mW})$ via \eqref{eqn:projectTangent}. \;
	Set the descent direction $\bar{\eta}_0 = -\mathrm{grad} \bar{f}(x_0) $. \;
	\Repeat{convergence.}{
		Select the step size $\alpha_k\in\bbR$ by the Wolfe condition and set
		$x_{k+1}=\sR_{x_k}(\alpha_k \cdot \bar{\eta}_k)$. \; 
		Compute $\bar{\zeta}_{k+1} = \sP_{x_k\to x_{k+1}}(\bar{\eta}_k)$. \; 
		Compute $\frac{\partial F}{\partial \mU}$ and $\frac{\partial F}{\partial  \mW}$ at the current $x_k = (\mU_{k}, \mW_{k})$. \;
		Compute the gradient $\mathrm{grad} \bar{f}(x_{k+1}) = \Psi_{x_{k+1}} (\frac{\partial F}{\partial \mU}, \frac{\partial F}{\partial  \mW})$ via \eqref{eqn:projectTangent}. \;
		Set  $\bar{\eta}_{k+1}=-\grad \bar{f}( x_{k+1})+
		\beta_{k+1}\bar{\zeta}_{k+1}$ with $\beta$ being specified by~\eqref{eqn:FRrule} or~\eqref{eqn:PRrule}. \;
		Set $k=k+1$. \;
	}
	Compute matrix product $\mD = \mU_{k}\mW_{k}\mU_{k}\trans$. \;
	Compute the eigendecomposition of $\mD = \mV\mE\mV\trans$. 
	Let the estimator of ${\mU}$ be the first $R$ columns of $\mV$. 
	Let the estimator of ${\mW}$ be the first $R$ diagonal elements of $\mE$. \;
	\caption{The Conjugate Gradient Algorithm over $\sM$. \label{algorithm:CG}}
\end{algorithm2e}

In Algorithm~\ref{algorithm:CG}, there are  five main steps (Line 6--10). 
In Line~6, we move the current point $x_{k}=(\mU_k,\mW_k)$ along the search direction $ \bar{\eta}_k$ and with the retraction operation $\sR_{x_k}$ in~\eqref{eqn:subRectraction}.  The step size  $\alpha_k$ is selected by the Wolfe condition (see \cite{nocedal2006numerical} and  its manifold counterpart which can be found in, for example, in \citep{ring2012optimization}). 
In Line~7, the search direction  $\bar{\eta}_k$  is transported onto the tangent space of the updated point  $x_{k+1}$. We calculate the  Euclidean derivatives  of the objective function $F$ in Line~8, and then project them onto the   manifold tangent space in Line~9. 
After that, in Line~10, a new search direction $ \bar{\eta}_{k+1}$  is computed as a combination of the translated search direction  $\bar{\zeta}_{k+1}$ and the calculated negative manifold gradient $-\grad \bar{f}( x_{k+1})$. The weights of this combination is based on the commonly used Fletcher-Reeves \citep{fletcher1964function} and the Polak-Ribi\`ere \citep{polak1971computational} rules. To be specific, we have
\begin{quote}
	\begin{enumerate}
		\item Fletcher-Reeves: 
		\begin{equation} \label{eqn:FRrule}
			\beta_{k+1}=\langle
			\grad \bar{f}(x_{k+1}), \grad \bar{f}  (x_{k+1})\rangle /
			\langle  \grad \bar{f} (x_k), \grad \bar{f} (x_k)\rangle;
		\end{equation}
		\item Polak-Ribi\`ere:
		\begin{equation} \label{eqn:PRrule}
			\beta_{k+1}=\langle
			\grad \bar{f} (x_{k+1}), \grad \bar{f} (x_{k+1})-\bar{\xi}_k\rangle /
			\langle \grad \bar{f} (x_k), \grad \bar{f} (x_k)\rangle 
		\end{equation}
		where $\bar{\xi}_k = \sP_{x_k\to x_{k+1}} (\grad \bar{f}(x_k)$).
	\end{enumerate}
\end{quote}

{The sequence of  directions $\bar{\eta}_0,\bar{\eta}_1,\bar{\eta}_2,\dots$
for updating $x = (\mU,\mW)$ resembles the classical conjugate gradient direction \citep{nocedal2006numerical}. For the classical conjugate gradient algorithm with the variable $x\in\bbR^p$ in a Euclidean space, suppose the (local) objective function $\bar{f}(x)$ has a quadratic form like $\bar{f}(x) = x\trans \mA x/2 + \vb\trans \vx$ for some positive definite matrix $\mA$ and vector $\vb$ of proper size. In this case, the classical conjugate gradient algorithm ensures the updating directions are conjugate to each other, i.e.,  $\bar{\eta}_i \trans \mA \bar{\eta}_j = 0$ for $i\neq j$. Moreover, the classical conjugate gradient method ensures that $x_{k+1}$ minimizes $\bar{f}(x)$
inside the subspace $x_0 + \text{span}(\bar{\eta}_0, \bar{\eta}_1,\cdots, \bar{\eta}_k )$, where $x_0$ is the initial value. 
As a notable feature of the classical conjugate gradient algorithm, the next direction $\bar{\xi}_{k+1}$ can be computed as a linear combination of the gradient at the current point $-\grad \bar{f}(x_{k+1})$ and the previous search direction $\bar{\xi}_k$. This is the same as Line~10 of Algorithm~\ref{algorithm:CG}, except that the previous search direction $\bar{\xi}_{k+1}$ is vector transported to the current point $x_{k+1}$ in Line~7. }


Finally, note that the working domain $\sM = \text{St}(R, K)\times \sS_{++}(R)$ 
of Algorithm~\ref{algorithm:CG} is larger than the constraint of~\eqref{eqn:firstOptimization}. In~\eqref{eqn:firstOptimization}, the matrix $\mW$ is restricted to be a diagonal matrix with positive elements. 
At the end (Line 13--14) of Algorithm~\ref{algorithm:CG}, we recover this structure. 
More precisely, let $({\mU}_{k},{\mW}_{k})\in\sM$ be the output of the iterative steps (Line~5--12) in Algorithm~\ref{algorithm:CG}.
We then compute the matrix product $\mD := {\mU}_{k}{\mW}_{k}{\mU}_{k}\trans$ and its eigendecomposition $\mD = \mV\mE\mV\trans$. 
The algorithm will output the first $R$ columns of $\mV$ and the first $R$ diagonal elements of $\mE$, respectively. 
The outputs correspond to the updated values of ${\mU}$ and ${\mW}$ when $\sigma_{e}^2$ is fixed in Algorithm \ref{alg:outerloop}. 

{
We conclude this section by remarking the computational complexities of the proposed CG method and the alternatives of EM and REML. Recall in our setting, $K$ is the number of basis functions, $R$ is the number of principal component functions (with $R \leq K$), and $M_n$ is the number of observation points for the $n$-th sample, $n=1,\dots, N$. 
For presentation simplicity, we assume $M_1, \dots, M_N$ are all equal to $M$.
The computational complexity of our proposed conjugate gradient algorithm (CG) includes two parts:
\begin{enumerate}[\hspace{18pt}]
	\item[C1] Computing the objective function and its Euclidean gradient requires a computational complexity of order $\mathcal{O}(N(R^3+MKR))$, after using the Cholesky decomposition, the matrix determinant lemma, and the Woodbury matrix identity for acceleration.
	\item[C2] For manifold-related operations, the computational complexity of calculating the gradient projection, vector transport, and retraction is  $\mathcal{O}(K^2R)$.
\end{enumerate}
The computational complexity in C1 also applies to EM (for evaluating the log-likelihood) and REML. The overhead in C2 for manifold-related computation is minor when the sample size $N$ is relatively large. }

\section{Numerical Studies} \label{sec:numerics}

In this section, our method is compared with several existing methods both on simulated datasets and a set of real astronomical data. The competing methods have the same loss function~\eqref{eqn:loss:logdet} as ours, but distinct optimization algorithms are employed. 
Our method minimizes the loss via the conjugate gradient algorithm over the product manifold in Section~\ref{sec:algorithm}, which will be denoted as CG in the following. 
The alternative methods include the reduced rank model~\citep{james2000principal} using the EM algorithm (denoted as EM) and the restricted maximal likelihood method~\citep[REML,][]{peng2009geometric}. 
The implementation of the alternative methods is based on the R package \texttt{fpca} \citep{fpcaPackage}.
The tuning parameters of all these methods are the number of spline knots, and they are tuned via cross-validation (CV).

\subsection{Simulation} \label{sec:numerics:simulation}

In the simulation, the random function are generated by a rank-$R_0$ model 
\begin{equation} \label{eqn:curves}
	x(u) = \sum_{r=1}^{R_0}\lambda_{0r}^{1/2}\theta_r\psi_{0r}(u),    
\end{equation}
where $\lambda_{0r}$'s are eigenvalues and $\theta_r$'s are random scores following standard normal distribution. The true eigenfunctions $\psi_{0r}$'s in~\eqref{eqn:curves} are constructed from sinusoid functions over $[0, 1]$ as in \cite{peng2009geometric}.

Similar to~\cite{peng2009geometric}, we consider two settings called \textit{easySin} and \textit{pracSin}. 
{
Generally, the \textit{easySin} setting has fewer number of eigenfunctions and each eigenfunction (as random combination of 5 sinusoid bases) has less variation within its domain. The eigenfunctions in this setting are relatively easy to be estimated. On the other hand, in the \textit{pracSin} setting, each eigenfunction (as a random combination of 10 sinusoid bases) has more variation in the domain. The \textit{pracSin} setting also has a slightly larger number of eigenfunctions and thus is closer to the real cases in practice. More precisely, in} the \textit{easySin} setting, the true rank is $R_0 = 3$  with eigenvalues: 1, 0.66, and 0.517. 
For the eigenfunctions, we consider $5$ sinusoid functions denoted as $\vf_{1}(u) = (\sin(u), \sin(2u), \ldots, \sin(5u))$. 
Though any three of these sinusoidal functions can be directly set as the true eigenfunctions, we decide to make the estimation task more interesting by creating linear combinations of them. 
To be specific, in each replication, we generate a $5 \times 3$ matrix $\mT^{(1)}$ whose elements are drawn from the standard normal distribution. Let $\mQ^{(1)}$ be the Q-factor of the QR decomposition of $\mT^{(1)}$, so that $\mQ^{(1)}$ is an orthonormal matrix. For $r=1,2,3$, we set the $r$-th true eigenfunction as $\psi_{0r}(u) = \vf_{1}^{\top}(u)\mQ_{:r}^{(1)}$ with $\mathbf{Q}_{:r}^{(1)}$ being the $r$-th column of $\mathbf{Q}^{(1)}$. 

The \textit{pracSin} setting is designed to be a more challenging setting, for which the rank is higher and the eigenfunctions have more complex shapes. 
Specifically, the true rank is set as $R_0 = 5$ with eigenvalues: 1, 0.66, 0.517, 0.435, and 0.381. 
Let $\vf_2(u) = (\sin(u), \sin(2u), \ldots, \sin(10u))$ and consider their linear combinations for eigenfunctions. We similarly denote $\mQ^{(2)}$ as the orthonormal Q-factor of the QR decomposition of some $10 \times 5$ random matrix $\mT^{(2)}$, whose elements are independently drawn from the standard normal distribution for different replicates. 
We set the $r$-th true function as $\psi_{0r}(u) = \vf_2^{\top} (u) \mQ^{(2)}_{:r}$ with $\mathbf{Q}_{:r}^{(2)}$ being the $r$-th column of $\mathbf{Q}^{(2)}$, $r=1,\ldots, 5$.

To generate $N$ sparse and noisy observations, we begin with the sparsely sampled curves $x_1,\cdots, x_N$.
In particular, the $n$-th sample curve $x_n$ has $M_n$ sparse observations, where $M_n$ is drawn from the discrete uniform distribution over $\{2,3,\cdots, 10\}$. For $j=1,\cdots, M_n$, the observation time points $u_{nj}$'s are sampled from the uniform distribution over $\sU =[0, 1]$.
The observed value at $u_{nj}$ is $y_{nj} = x(u_{nj}) + \sigma_e\epsilon_{nj}$ with $\sigma_e = 1/4$. The noise $\epsilon_{nj}$ follows one of the three distributions: the standard normal $\mathcal{N}(0, 1)$, the rescaled student t distribution $t_3/\sqrt{3}$ with $3$ degrees of freedom, and the uniform distribution $U[-\sqrt{3}, \sqrt{3}]$.
In this way, a collection of sparse and noisy observations $\{(u_{nj}, y_{nj})\}_{j=1}^{M_n}$ are generated under each setting, $n =1, \dots, N$.
For \textit{easySin}, $N = 50$ samples are generated; as for \textit{pracSin}, we generate $100$ and $500$ samples. 
In each setting, $500$ replications are carried out.

\begin{table}[h!]
	\caption{Simulation result for the \textit{easySin} setting.  The reported values ($\times 10^{-1}$) are the average estimation errors for each eigenfunction $\|\hat{\psi}_r - \psi_{0r}\|_{L_2}$ based on 500 replications.  The reported numbers are the actual values multiplied by 10. The numbers in the parentheses are the standard errors. The first row indicates the initialization methods of EM, LS, and LOC. The row `\# Conv' is the number of converged replicates.}
	\label{tab:easy_sin_500}
	\vspace{3pt}
	\centering
	\resizebox{\textwidth}{!}{
		\begin{tabular}{ c| c c |  c c c |  c c c c}
			\hline
			& \multicolumn{2}{c}{EM ($\times 10^{-1}$)} & \multicolumn{3}{|c|}{LS  ($\times 10^{-1}$)} & \multicolumn{3}{c}{LOC ($\times 10^{-1}$)}\\
			\hline  \hline
			\multicolumn{9}{c}{$N = 50$} \\
			\hline 
			$\varepsilon \sim \text{Normal}$ & 
			REML & CG  & REML & EM & CG   & REML & EM & CG \\
			\hline
			$  \hat{\psi}_{1} $ & 4.57(0.14) & \textbf{4.42(0.12)} & 7.25(0.18) & 4.91(0.14) & \textbf{4.42(0.12)} & 6.92(0.18) & 4.90(0.13) & \textbf{4.42(0.12)}\\ 
			$  \hat{\psi}_{2} $ & 7.20(0.17) & \textbf{7.09(0.16)} & 9.42(0.17) & 7.45(0.16) & \textbf{7.09(0.16)} & 8.88(0.17) & 7.47(0.16) & \textbf{7.06(0.16)}\\ 
			$  \hat{\psi}_{3} $ & 6.27(0.16) & \textbf{6.24(0.15)} & 9.86(0.18) & 6.79(0.17) & \textbf{6.25(0.15)} & 9.07(0.18) & 6.87(0.17) & \textbf{6.23(0.15)}\\ 
			\hline
			\# Conv   & 482 & 500 & 362 & 500 & 500 & 399 & 500 & 500 \\
			\hline
			$\varepsilon \sim t_3$ & \\
			\hline
			$  \hat{\psi}_{1} $ & 4.87(0.14) & \textbf{4.82(0.13)} & 7.45(0.19) & 5.22(0.14) & \textbf{4.81(0.13)} & 7.00(0.18) & 5.22(0.14) & \textbf{4.82(0.13)}\\ 
			$  \hat{\psi}_{2} $ & \textbf{7.01(0.16)} & 7.05(0.15) & 9.15(0.17) & 7.46(0.16) & \textbf{7.03(0.15)} & 8.72(0.16) & 7.53(0.16) & \textbf{7.06(0.15)}\\ 
			$  \hat{\psi}_{3} $ & 6.03(0.15) & \textbf{5.91(0.15)} & 9.65(0.19) & 6.59(0.16) & \textbf{5.91(0.15)} & 8.85(0.18) & 6.62(0.16) & \textbf{5.93(0.15)}\\ 
			\hline
			\# Conv  & 474 & 500 & 365 & 500 & 500 & 406 & 500 & 500 \\
			\hline
			
			$\varepsilon \sim \text{Unif}$ & \\
			\hline
			$ \hat{\psi}_{1} $ & 4.86(0.15) & \textbf{4.74(0.14)} & 7.43(0.18) & 5.33(0.15) & \textbf{4.75(0.14)} & 7.02(0.18) & 5.27(0.15) & \textbf{4.76(0.14)}\\ 
			$  \hat{\psi}_{2} $ & 7.30(0.17) & \textbf{7.22(0.16)} & 9.32(0.16) & 7.66(0.16) & \textbf{7.22(0.16)} & 9.10(0.17) & 7.73(0.16) & \textbf{7.21(0.16)}\\ 
			$  \hat{\psi}_{3} $ & 6.03(0.16) & \textbf{6.01(0.15)} & 9.68(0.19) & 6.69(0.16) & \textbf{6.01(0.15)} & 9.04(0.18) & 6.89(0.17) & \textbf{5.99(0.15)}\\ 
			\hline
			\# Conv   & 480 & 500 & 398 & 500 & 500 & 378 & 500 & 500  \\
			\hline
			\hline
		\end{tabular}
	}
\end{table}

The functional principal components are then estimated from the simulated datasets by the proposed CG method and competitive EM and REML. 
For fitting, cubic spline basis (with order $4$) is used. The number of knots is selected among $\{4,5,\ldots,12\}$  for the \textit{easySin} setting, and from $\{8,9,\ldots,15\}$ for the \textit{pracSin} setting. The knot selection is conducted by ten-fold CV, and the loss function used in CV is the same as \eqref{eqn:loss:logdet}. 
As in \cite{peng2009geometric}, we fix the number of principal functions at the true one $R = R_0$ for all methods in the simulation studies. 
Lastly, to evaluate the performance of various methods, we compute the average estimation errors in $L_2$ norm, i.e., $\|\hat{\psi}_r - \psi_{0r}\|_{L_2}$, for each eigenfunction.

To compare the stability of various methods, we also consider three different types of initial values: LOC, LS, and EM.
Specifically, (1) LOC: these initial values are the estimation results of the local polynomial method~\citep[LOC,][]{yao05}. LOC fits the mean and covariance functions through nonparametric local regression and thus itself does not require initial values. (2) LS: these initial values are obtained by a least squares procedure. It fits the linear model $\vy_n = \mB_{n} \boldsymbol{\gamma}_n + \vepsilon_{n}$ to each sample independently. The initial value of $\mU$ is then extracted from the left singular vector matrix of $\mGamma = (\hat{\boldsymbol{\gamma}}_1, \ldots, \hat{\boldsymbol{\gamma}}_N)$ where   $\hat{\boldsymbol{\gamma}}_n = (\mB_n\trans\mB_n)^{-1}\mB_n\trans\vy_n $.
(3) EM: the final estimate from the EM method~\cite{james2000principal} is directly used. This initial value is suggested in \cite{peng2009geometric} for REML. We thus compare the performance of REML and CG under this initialization.

\begin{table}[h!]
	\caption{Simulation result for the \textit{pracSin} setting. The reported values ($\times 10^{-1}$) are the average estimation errors for each eigenfunction
		$\|\hat{\psi}_r - \psi_{0r}\|_{L_2}$  based on 500 replications.  The reported numbers are the actual values multiplied by 10.
		The numbers in the parentheses are the standard errors. The first row indicates the initialization methods of EM, LS, and LOC. The row `\# Conv' is the number of converged replicates.}
	\label{tab:prac_Sin}
	\vspace{3pt}
		\centering
		\resizebox{\textwidth}{!}{
			\begin{tabular}{ c |  c c | c c c  |   c c c}
				\hline
	 & \multicolumn{2}{|c|}{EM ($\times 10^{-1}$)} & \multicolumn{3}{|c|}{LS ($\times 10^{-1}$)}  & \multicolumn{3}{c}{LOC ($\times 10^{-1}$)} \\
					 \hline \hline
				 				\multicolumn{9}{c}{$N =100$  }   \\
				 	\hline				
				$ \varepsilon \sim \text{Normal}$ & REML & CG   & REML & EM & CG  & REML & EM & CG \\
				\hline
				$\hat{\psi}_{1}$ &   6.80(0.15) & \textbf{ 4.53(0.10)} & 11.39(0.11) &   5.00(0.12) & \textbf{ 4.56(0.10)} & 11.72(0.11) &   5.17(0.12) & \textbf{ 4.60(0.10)}\\ 
				$\hat{\psi}_{2}$ &   9.32(0.14) & \textbf{ 7.59(0.13)} & 12.19(0.08) &   8.06(0.14) & \textbf{ 7.62(0.13)} & 12.16(0.08) &   8.12(0.14) & \textbf{ 7.67(0.13)}\\ 
				$\hat{\psi}_{3}$ & 10.23(0.12) & \textbf{ 9.20(0.14)} & 12.51(0.06) &   9.64(0.13) & \textbf{ 9.18(0.13)} & 12.28(0.07) &   9.59(0.13) & \textbf{ 9.24(0.13)}\\ 
				$\hat{\psi}_{4}$ & 10.37(0.12) & \textbf{ 9.61(0.13)} & 12.65(0.06) &   9.76(0.13) & \textbf{ 9.67(0.13)} & 12.54(0.07) &   9.92(0.13) & \textbf{ 9.65(0.13)}\\ 
				$\hat{\psi}_{5}$ & 10.24(0.13) & \textbf{ 8.17(0.14)} & 12.50(0.06) &   8.69(0.15) & \textbf{ 8.24(0.14)} & 12.51(0.07) &   8.96(0.15) & \textbf{ 8.28(0.14)}\\ 
				\hline
				\# Conv & 434 & 500 & 62 & 500 & 500 & 66 & 499 & 500 \\
				\hline
				$ \varepsilon \sim t_3$ & \\
				\hline
				$\hat{\psi}_{1}$ &   6.90(0.15) & \textbf{ 4.74(0.10)} & 11.41(0.12) &   5.26(0.12) & \textbf{ 4.77(0.10)} & 11.42(0.11) &   5.34(0.13) & \textbf{ 4.82(0.11)}\\ 
				$\hat{\psi}_{2}$ &   9.41(0.13) & \textbf{ 7.60(0.14)} & 12.23(0.08) &   7.93(0.14) & \textbf{ 7.67(0.14)} & 12.14(0.08) &   7.94(0.14) & \textbf{ 7.68(0.14)}\\ 
				$\hat{\psi}_{3}$ & 10.39(0.12) & \textbf{ 9.33(0.13)} & 12.37(0.07) &   9.54(0.13) & \textbf{ 9.41(0.13)} & 12.29(0.07) &   9.55(0.13) & \textbf{ 9.44(0.13)}\\ 
				$\hat{\psi}_{4}$ & 10.52(0.12) & \textbf{ 9.51(0.13)} & 12.56(0.06) &   9.83(0.13) & \textbf{ 9.53(0.13)} & 12.48(0.06) &   9.90(0.13) & \textbf{ 9.62(0.13)}\\ 
				$\hat{\psi}_{5}$ & 10.43(0.13) & \textbf{ 8.48(0.14)} & 12.54(0.06) &   8.82(0.14) & \textbf{ 8.52(0.14)} & 12.54(0.06) &   8.94(0.15) & \textbf{ 8.56(0.14)}\\ 
				\hline
				\# Conv & 437 & 500 & 57 & 500 & 500 & 77 & 500 & 500 \\
				\hline
				$ \varepsilon \sim \text{Unif}$ & \\
				\hline
				$\hat{\psi}_{1}$ &   6.70(0.15) & \textbf{ 4.67(0.11)} & 11.46(0.11) &   5.15(0.13) & \textbf{ 4.64(0.11)} & 11.50(0.11) &   5.29(0.13) & \textbf{ 4.66(0.11)}\\ 
				$\hat{\psi}_{2}$ &   9.07(0.14) & \textbf{ 7.60(0.14)} & 12.18(0.08) &   7.91(0.14) & \textbf{ 7.59(0.14)} & 12.18(0.08) &   7.99(0.13) & \textbf{ 7.62(0.13)}\\ 
				$\hat{\psi}_{3}$ & 10.16(0.12) & \textbf{ 9.20(0.14)} & 12.47(0.07) &   9.54(0.13) & \textbf{ 9.20(0.13)} & 12.20(0.08) &   9.69(0.13) & \textbf{ 9.20(0.13)}\\ 
				$\hat{\psi}_{4}$ & 10.17(0.13) & \textbf{ 9.56(0.14)} & 12.54(0.06) &   9.85(0.13) & \textbf{ 9.54(0.14)} & 12.35(0.07) &   9.89(0.13) & \textbf{ 9.56(0.14)}\\ 
				$\hat{\psi}_{5}$ & 10.35(0.13) & \textbf{ 8.41(0.14)} & 12.42(0.06) &   8.86(0.14) & \textbf{ 8.40(0.14)} & 12.52(0.06) &   8.90(0.14) & \textbf{ 8.47(0.14)}\\ 
				\hline
				\# Conv  & 451 & 500 & 59 & 500 & 500 & 73 & 496 & 500 \\
				\hline
				\hline 
				\multicolumn{9}{c}{$N =500$  }   \\
				\hline
				$ \varepsilon \sim \text{Normal}$ &   REML & CG   & REML & EM & CG  & REML & EM & CG\\
				\hline
				$\hat{\psi}_{1}$ & \textbf{ 1.88(0.03)} &   1.90(0.03) &   6.71(0.22) &   2.67(0.10) & \textbf{ 1.90(0.03)} &   8.04(0.22) &   2.93(0.10) & \textbf{ 1.90(0.03)}\\ 
				$\hat{\psi}_{2}$ & \textbf{ 3.27(0.07)} &   3.28(0.07) &   7.95(0.19) &   4.20(0.13) & \textbf{ 3.28(0.07)} &   9.17(0.19) &   4.48(0.13) & \textbf{ 3.28(0.07)}\\ 
				$\hat{\psi}_{3}$ &   5.03(0.12) & \textbf{ 5.01(0.12)} &   9.06(0.18) &   5.88(0.15) & \textbf{ 5.01(0.12)} &   9.86(0.16) &   6.01(0.14) & \textbf{ 5.01(0.11)}\\ 
				$\hat{\psi}_{4}$ & \textbf{ 6.12(0.14)} &   6.17(0.14) &   9.48(0.17) &   6.98(0.16) & \textbf{ 6.18(0.14)} & 10.35(0.15) &   7.03(0.15) & \textbf{ 6.18(0.14)}\\ 
				$\hat{\psi}_{5}$ & \textbf{ 4.89(0.13)} &   4.94(0.13) &   8.97(0.19) &   6.05(0.17) & \textbf{ 4.95(0.13)} &   9.88(0.17) &   6.07(0.17) & \textbf{ 4.96(0.13)}\\ 
				\hline
				\# Conv & 486 & 500 & 248 & 500 & 500 & 221 & 500 & 500\\
				\hline
				$ \varepsilon \sim t_3 $ \\
				\hline
				$\hat{\psi}_{1}$ & \textbf{ 1.87(0.03)} &   1.89(0.03) &   7.26(0.22) &   2.60(0.10) & \textbf{ 1.89(0.03)} &   7.98(0.21) &   2.85(0.10) & \textbf{ 1.89(0.03)}\\ 
				$\hat{\psi}_{2}$ &   3.38(0.08) & \textbf{ 3.36(0.07)} &   8.43(0.20) &   4.22(0.12) & \textbf{ 3.36(0.07)} &   9.23(0.19) &   4.54(0.12) & \textbf{ 3.36(0.07)}\\ 
				$\hat{\psi}_{3}$ &   5.14(0.12) & \textbf{ 5.11(0.12)} &   9.30(0.18) &   5.85(0.14) & \textbf{ 5.07(0.12)} & 10.09(0.16) &   6.14(0.14) & \textbf{ 5.08(0.12)}\\ 
				$\hat{\psi}_{4}$ &   6.21(0.14) & \textbf{ 6.10(0.14)} &   9.97(0.16) &   6.81(0.15) & \textbf{ 6.06(0.14)} & 10.39(0.15) &   7.02(0.16) & \textbf{ 6.07(0.14)}\\ 
				$\hat{\psi}_{5}$ &   5.01(0.13) & \textbf{ 4.94(0.13)} &   9.33(0.18) &   5.86(0.17) & \textbf{ 4.92(0.13)} & 10.15(0.16) &   5.99(0.17) & \textbf{ 4.93(0.13)}\\ 
				\hline
				\# Conv & 491 & 500 & 245 & 500 & 500 & 224 & 500 & 500 \\
				\hline
				$ \varepsilon \sim \text{Unif}$ & \\
				\hline
				$\hat{\psi}_{1}$ & \textbf{ 1.86(0.03)} &   1.88(0.03) &   6.78(0.22) &   2.67(0.10) & \textbf{ 1.88(0.03)} &   8.08(0.22) &   2.89(0.10) & \textbf{ 1.88(0.03)}\\ 
				$\hat{\psi}_{2}$ & \textbf{ 3.31(0.08)} &   3.32(0.07) &   8.11(0.20) &   4.18(0.13) & \textbf{ 3.32(0.07)} &   9.13(0.19) &   4.47(0.13) & \textbf{ 3.32(0.07)}\\ 
				$\hat{\psi}_{3}$ &   5.09(0.12) & \textbf{ 5.01(0.12)} &   8.99(0.18) &   5.77(0.15) & \textbf{ 5.01(0.12)} &   9.90(0.16) &   6.03(0.14) & \textbf{ 5.02(0.12)}\\ 
				$\hat{\psi}_{4}$ &   6.21(0.14) & \textbf{ 6.18(0.14)} &   9.53(0.17) &   6.81(0.16) & \textbf{ 6.17(0.14)} & 10.31(0.15) &   6.98(0.16) & \textbf{ 6.18(0.14)}\\ 
				$\hat{\psi}_{5}$ & \textbf{ 4.91(0.13)} &   5.00(0.13) &   8.81(0.19) &   5.94(0.17) & \textbf{ 4.99(0.13)} & 10.07(0.17) &   6.04(0.17) & \textbf{ 5.00(0.13)}\\ 
				\hline
				\# Conv & 478 & 500 & 260 & 500 & 500 & 217 & 500 & 500 \\
				\hline
				\hline
			\end{tabular}
		}
	\end{table}

	Tables~\ref{tab:easy_sin_500} and \ref{tab:prac_Sin} summarize the results for the settings of \textit{easySin} and \textit{pracSin}, respectively. All the reported numbers are the actual values multiplied by 10. Three groups of columns correspond to distinct initial values (from EM, LS, and LOC). 
	The results of each distribution type of the noise $\epsilon$ (Gaussian, t-distribution, and uniform) are grouped into blocks. In the table, we report the average estimation error $\|\hat{\psi}_r - \psi_{0r}\|_{L_2}$ for each eigenfunction, as well as the number of converged replicates (the row denoted `\# Conv') for each method. The estimation errors are computed and averaged only among the converged replicates of each method. 
	
	Table~\ref{tab:easy_sin_500} shows that the performance of 
	the proposed CG is stable and reasonably better than the competitors.  	
	Given EM as initial values, the performances of REML and CG are very close generally. 
	However, if the initial values are taken other than EM, the performance of REML deteriorates considerably in the sense that its estimation error is much larger when LS or LOC is used for initialization. 
	Moreover, REML has 4\%--30\% un-converged replicates out of the total 500, especially under the initialization method of LS or LOC.
	Note that in this study, we use the converged replicates to compute the estimation errors; the performance of REML may be much worse when a practitioner uses the last iteration as the output in real applications.
	This phenomenon is consistent with statements in \cite{peng2009geometric} that REML is very sensitive to the choice of initial values and the unconvergence of REML mainly comes from unsuitable initial estimates.
	Unlike REML, the performance of our proposed CG method is stable across different initials with the highest accuracy for eigenfunction estimation in almost all cases. The improvement of CG over REML illustrates the benefit of the proposed conjugate gradient algorithm. 
	The results in Table \ref{tab:prac_Sin} are generally similar as those in Table~\ref{tab:easy_sin_500}.
	Since the true eigenfunctions have complex shapes under the \textit{pracSin} setting, the sample size $N=100$ is not sufficient for REML to work well.
	To be specific, with LS and LOC as initial values, REML has less than 100 converged replicates out of the total 500, and its estimation error is much larger than others. 
	Our proposed CG outperforms the competitors and is more stable than REML and EM with different initial values. 
	When the sample size increases to 500, the performance of all methods gets improved. However, CG remains superior to the others.
	
	
	
	
{
To compare the running efficiency between our algorithm with the traditional methods of EM and REML, we run all methods on the same computation platform with 2.40 GHz Intel (R) Xeon (R) E5-2680 v4 28-Core CPU using R. 
We record the CPU running time of each method for all settings to compare their computational efficiency. 
We depict the boxplots of the running time (in seconds) of each method among 500 repetitions with normal distributed noises. 
Figures~\ref{fig:boxplot_time_easySin}--\ref{fig:boxplot_time_pracSin2} correspond to the \textit{easySin} setting with $N=50$, the \textit{pracSin} setting with $N=100$, and the \textit{pracSin} setting with $N=500$, respectively. 
The results of the running time when the noises are rescaled student t distributed and uniform distributed are similar. 
These plots evidently show that the running time of the proposed CG method is less than REML and EM on average. For example, the average running time of the REML is above $30$ seconds for the \textit{pracSin} setting with $N=500$; whereas the proposed CG takes less then $3$ seconds on average for the same setting. 
\begin{figure}[h!]
    \centering
    \includegraphics[width = 0.7\textwidth]{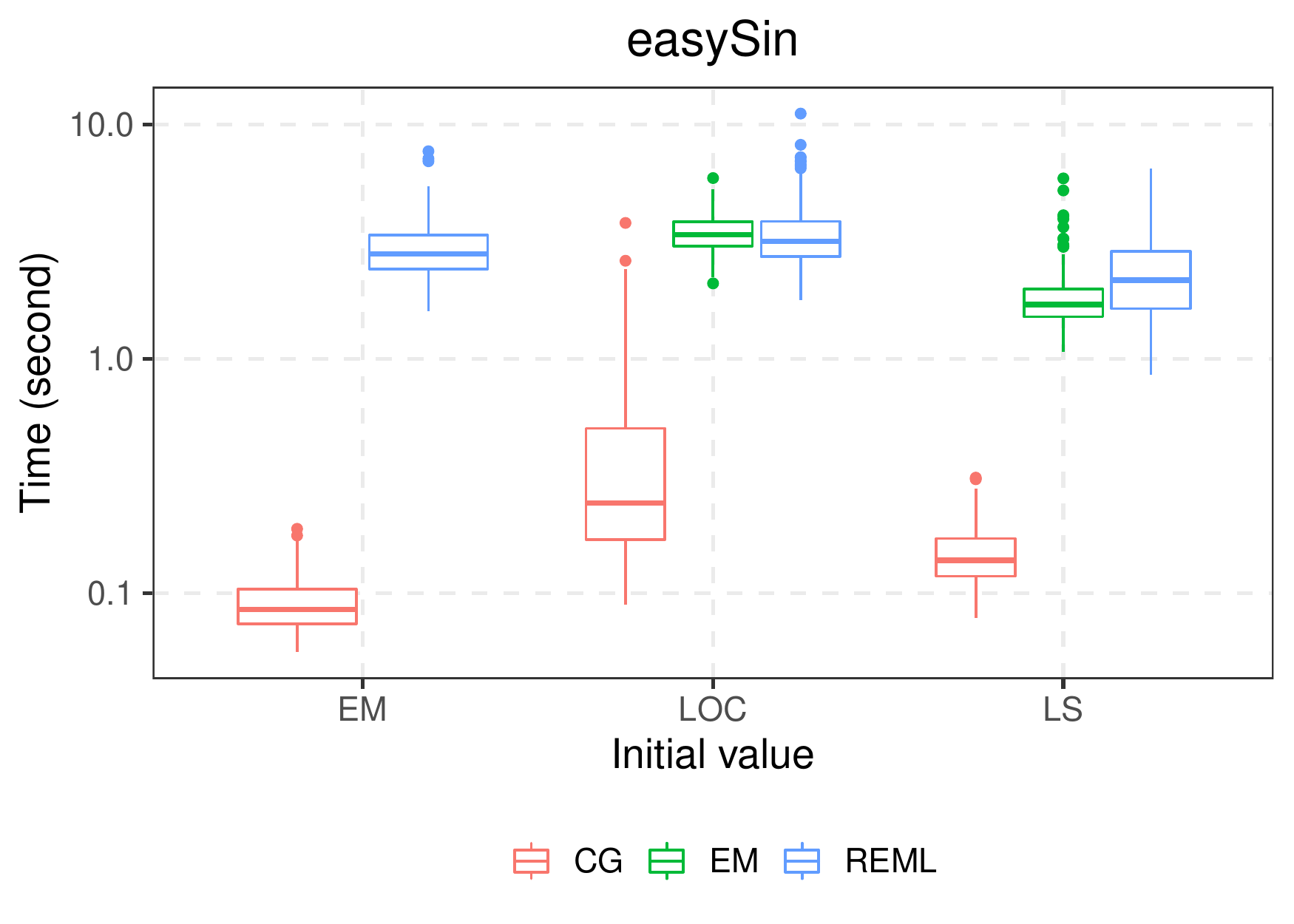}
	\caption{The boxplot of running time of each method with different initial values for the \textit{easySin} setting under Gaussian noises. The results are based on $500$ simulation replications, and the $y$-axis is plotted in $\log_{10}$ scale.  
	}
	\label{fig:boxplot_time_easySin}
\end{figure}
\begin{figure}[h!]
	\centering
	\includegraphics[width = 0.7\textwidth]{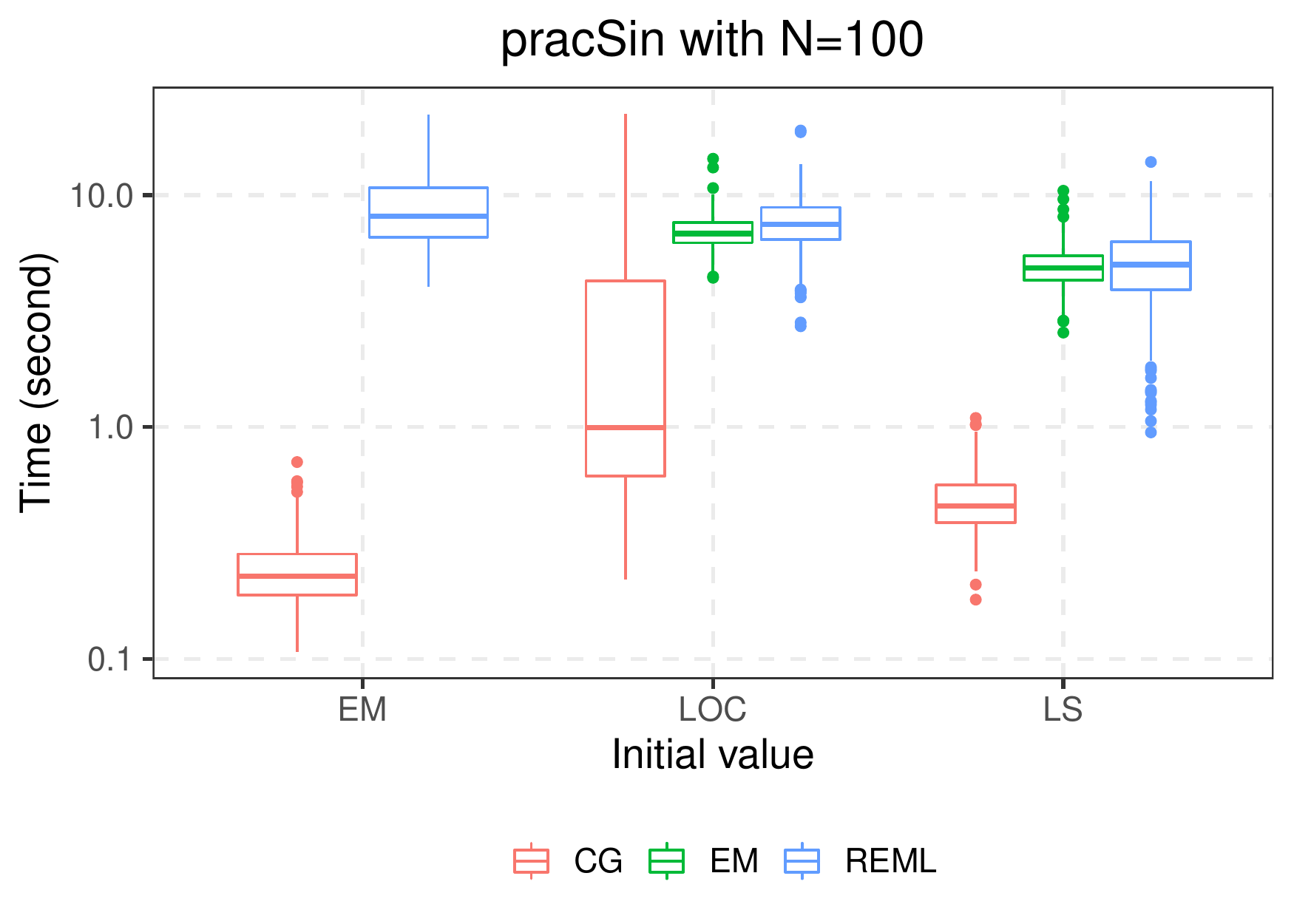}
	\caption{The boxplot of running time of each method with different initial values for the \textit{pracSin} setting of sample size $N = 100$ under Gaussian noises. The results are based on $500$ simulation replications, and the $y$-axis is plotted in $\log_{10}$ scale. 
	}\label{fig:boxplot_time_pracSin1}
\end{figure}
\begin{figure}[h!]
	\centering
	\includegraphics[width = 0.7\textwidth]{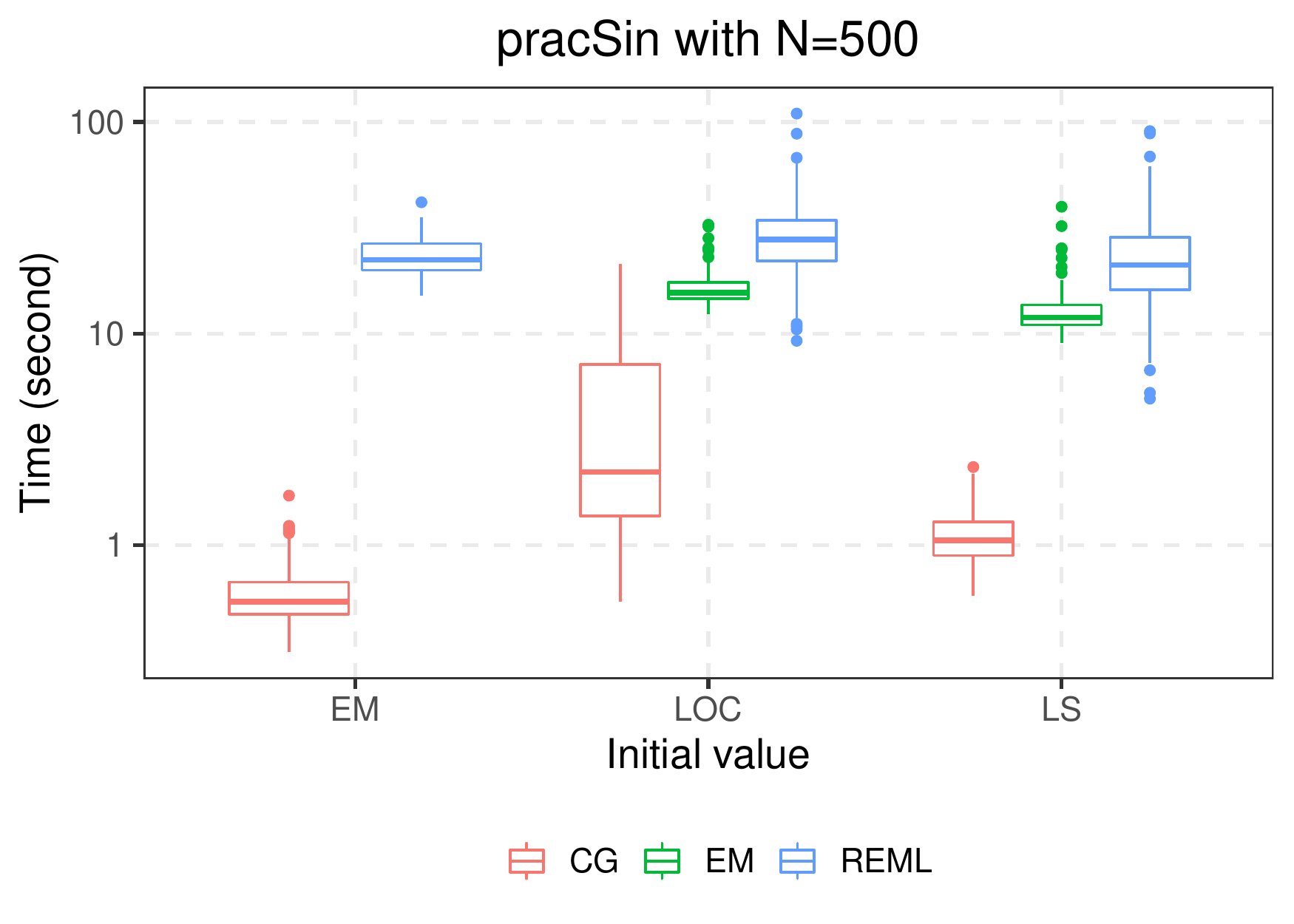}
	\caption{The boxplot of running time of each method with different initial values for the \textit{pracSin} setting of sample size $N = 500$ under Gaussian noises. The results are based on $500$ simulation replications, and the $y$-axis is plotted in $\log_{10}$ scale. }
	\label{fig:boxplot_time_pracSin2}
\end{figure}	
}
	
	We also explore two more settings used in~\cite{peng2009geometric} which particularly favor REML. 
	The two settings are denoted as \textit{easy} and \textit{prac}. For the former, there are $3$ eigenvalues: 1, 0.66, and 0.517; and the eigenfunctions are directly generated by the cubic B-spline with $5$ equally spaced knots. The latter is more complex with $10$ equally spaced knots and $5$ eigenvalues: 1, 0.66, 0.517, 0.435, and 0.381. 
	The noises also follow Gaussian, rescaled student t, and uniform distributions.
	For both settings, we fix $R = R_0$ for all methods. 
	The results for these two settings are presented in Tables~\ref{tab:easy_data} and~\ref{tab:prac_data}, respectively. 
	It shows that the performance of REML becomes competitive under these settings when the output of EM is used as the initial value, and the sample size is large. However, even in these cases, the performance of the proposed CG is close to that of REML. For other cases, the CG remains to outperform REML and EM in terms of estimation accuracy and stability of initials.

	\begin{table}[h!]
		\caption{Simulation result for the \textit{easy} setting. The reported values ($\times 10^{-1}$) are the average estimation errors for each eigenfunction
			$\|\hat{\psi}_r - \psi_{0r}\|_{L_2}$  based on 500 replications.  The reported numbers are the actual values multiplied by 10.	The numbers in the parentheses are the standard errors. The first row indicates the initialization methods of EM, LS, and LOC. The row `\# Conv' is the number of converged replicates.}
		\label{tab:easy_data}
		\vspace{3pt}
		\begin{subtable}{\textwidth}
			\centering
			\resizebox{\textwidth}{!}{
				\begin{tabular}{ c| c c |  c c c |  c c c c}
					\hline
				   & \multicolumn{2}{c}{EM  ($\times 10^{-1}$) } & \multicolumn{3}{|c|}{LS  ($\times 10^{-1}$) } & \multicolumn{3}{c}{LOC  ($\times 10^{-1}$) }   \\
					 \hline \hline
					\multicolumn{9}{c}{$N =50$}\\
					\hline
$ \varepsilon \sim \text{Normal}$ & REML & CG  & REML & EM & CG   & REML & EM & CG	\\
					\hline
					$ \hat{\psi}_{1}$ & \textbf{5.06(0.14)} & 5.07(0.14) & 5.36(0.14) & 5.63(0.14) & \textbf{5.07(0.14)} & 6.00(0.16) & 5.67(0.14) & \textbf{5.07(0.14)}\\ 
					$ \hat{\psi}_{2}$ & \textbf{6.96(0.16)} & 7.06(0.15) & 7.12(0.16) & 7.26(0.15) & \textbf{7.05(0.16)} & 7.43(0.16) & 7.17(0.16) & \textbf{7.06(0.16)}\\ 
					$ \hat{\psi}_{3}$ & \textbf{5.56(0.14)} & 5.81(0.14) & 6.08(0.15) & 6.07(0.14) & \textbf{5.80(0.14)} & 6.87(0.18) & 6.06(0.15) & \textbf{5.81(0.14)}\\ 
					\hline
					\# Conv    & 494 & 500 & 475 & 500 & 500 & 463 & 500 & 500 \\
					\hline
					$ \varepsilon \sim t_3$ & \\
					\hline
					$ \hat{\psi}_{1}$ & 5.17(0.14) & \textbf{5.14(0.14)} & 5.55(0.15) & 5.77(0.14) & \textbf{5.15(0.14)} & 6.09(0.16) & 5.75(0.14) & \textbf{5.14(0.14)}\\ 
					$ \hat{\psi}_{2}$ & 6.95(0.16) & \textbf{6.86(0.16)} & 7.13(0.16) & 7.22(0.16) & \textbf{6.81(0.16)} & 7.62(0.16) & 7.16(0.16) & \textbf{6.84(0.16)}\\ 
					$ \hat{\psi}_{3}$ & \textbf{5.50(0.14)} & 5.67(0.15) & 6.02(0.15) & 5.96(0.15) & \textbf{5.61(0.15)} & 6.85(0.17) & 5.98(0.15) & \textbf{5.64(0.15)}\\ 
					\hline
					\# Conv   & 495 & 500 & 474 & 500 & 500 & 453 & 500 & 500  \\
					\hline
					$ \varepsilon \sim \text{Unif}$ & \\
					\hline
					$ \hat{\psi}_{1}$ & \textbf{5.19(0.15)} & 5.20(0.15) & 5.48(0.15) & 5.79(0.15) & \textbf{5.19(0.15)} & 5.79(0.16) & 5.87(0.15) & \textbf{5.19(0.15)}\\ 
					$ \hat{\psi}_{2}$ & \textbf{6.89(0.16)} & 7.01(0.16) & 7.03(0.16) & 7.15(0.16) & \textbf{6.97(0.16)} & 7.22(0.16) & 7.27(0.16) & \textbf{7.00(0.16)}\\ 
					$ \hat{\psi}_{3}$ & \textbf{5.26(0.14)} & 5.57(0.14) & 5.80(0.15) & 5.89(0.14) & \textbf{5.54(0.14)} & 6.64(0.17) & 5.90(0.14) & \textbf{5.57(0.14)}\\ 
					\hline
					\# Conv  & 492 & 500 & 476 & 500 & 500 & 461 & 500 & 500 \\
					\hline
					\hline
				\end{tabular}
			}
		\end{subtable}
	\end{table}

	\begin{table}[h!]
		\caption{Simulation result for the \textit{prac} setting.  The reported values ($\times 10^{-1}$) are the average estimation errors for each eigenfunction
			$\|\hat{\psi}_r - \psi_{0r}\|_{L_2}$  based on 500 replications.  The reported values are the actual values multiplied by 10.	The numbers in the parentheses are the standard errors. The first row indicates the initialization methods of EM, LS, and LOC. The row `\# Conv' is the number of converged replicates. }
		\label{tab:prac_data}
		\vspace{3pt}
			\centering
			\resizebox{\textwidth}{!}{
				\begin{tabular}{ c |  c c | c c c  |   c c c}
					\hline
					  & \multicolumn{2}{|c|}{EM ($\times 10^{-1}$)} & \multicolumn{3}{|c|}{LS ($\times 10^{-1}$)} & \multicolumn{3}{c}{LOC ($\times 10^{-1}$)}\\
					  \hline \hline
						 \multicolumn{9}{c}{$N =100$  } \\
					\hline 
					$ \varepsilon \sim \text{Normal}$ & REML & CG   & REML & EM & CG  & REML & EM & CG \\
					\hline
					$ \hat{\psi}_{1}$ & 7.37(0.19) & \textbf{4.88(0.12)} & 11.40(0.10) & 6.07(0.15) & \textbf{4.83(0.11)} & 11.68(0.10) & 6.30(0.15) & \textbf{4.92(0.12)}\\ 
					$ \hat{\psi}_{2}$ & 9.20(0.15) & \textbf{7.73(0.14)} & 12.39(0.07) & 8.38(0.14) & \textbf{7.69(0.13)} & 12.45(0.07) & 8.62(0.14) & \textbf{7.78(0.14)}\\ 
					$ \hat{\psi}_{3}$ & 10.28(0.13) & \textbf{9.51(0.13)} & 12.09(0.07) & 9.92(0.12) & \textbf{9.50(0.13)} & 12.08(0.07) & 10.08(0.12) & \textbf{9.65(0.13)}\\ 
					$ \hat{\psi}_{4}$ & 10.13(0.13) & \textbf{9.39(0.13)} & 12.03(0.08) & 9.56(0.13) & \textbf{9.43(0.13)} & 12.11(0.07) & 9.88(0.13) & \textbf{9.51(0.13)}\\ 
					$ \hat{\psi}_{5}$ & 9.71(0.15) & \textbf{8.52(0.14)} & 12.20(0.07) & 9.01(0.15) & \textbf{8.63(0.15)} & 12.18(0.07) & 9.52(0.14) & \textbf{8.88(0.15)}\\ 
					\hline
					\# Conv & 403 & 500 & 38 & 495 & 500 & 32 & 484 & 500 \\
					\hline
					$ \varepsilon \sim t_3$ & \\
					\hline
					$ \hat{\psi}_{1}$ & 7.40(0.19) & \textbf{4.85(0.12)} & 11.32(0.11) & 5.79(0.15) & \textbf{4.82(0.12)} & 11.66(0.10) & 6.09(0.15) & \textbf{4.95(0.12)}\\ 
					$ \hat{\psi}_{2}$ & 9.46(0.16) & \textbf{7.82(0.14)} & 12.24(0.08) & 8.36(0.15) & \textbf{7.78(0.14)} & 12.41(0.07) & 8.60(0.14) & \textbf{7.87(0.14)}\\ 
					$ \hat{\psi}_{3}$ & 10.12(0.13) & \textbf{9.45(0.13)} & 12.20(0.07) & 9.92(0.13) & \textbf{9.49(0.13)} & 12.17(0.07) & 10.02(0.12) & \textbf{9.56(0.13)}\\ 
					$ \hat{\psi}_{4}$ & 9.86(0.13) & \textbf{9.25(0.14)} & 12.05(0.07) & 9.49(0.14) & \textbf{9.21(0.14)} & 12.25(0.07) & 9.77(0.13) & \textbf{9.18(0.14)}\\ 
					$ \hat{\psi}_{5}$ & 9.53(0.15) & \textbf{8.47(0.15)} & 12.20(0.07) & 8.89(0.14) & \textbf{8.49(0.15)} & 12.23(0.07) & 9.43(0.14) & \textbf{8.65(0.15)}\\ 
					\hline
					\# Conv & 403 & 500 & 41 & 499 & 500 & 21 & 482 & 500\\
					\hline
					$ \varepsilon \sim \text{Unif}$ & \\
					\hline
					$ \hat{\psi}_{1}$ & 7.02(0.19) & \textbf{4.74(0.11)} & 11.67(0.10) & 5.87(0.15) & \textbf{4.73(0.11)} & 11.55(0.10) & 6.24(0.16) & \textbf{4.81(0.12)}\\ 
					$ \hat{\psi}_{2}$ & 9.10(0.15) & \textbf{7.66(0.13)} & 12.37(0.07) & 8.25(0.14) & \textbf{7.67(0.13)} & 12.32(0.07) & 8.55(0.14) & \textbf{7.72(0.13)}\\ 
					$ \hat{\psi}_{3}$ & 9.93(0.13) & \textbf{9.47(0.13)} & 12.03(0.07) & 9.67(0.12) & \textbf{9.56(0.13)} & 12.09(0.07) & 9.76(0.12) & \textbf{9.53(0.13)}\\ 
					$ \hat{\psi}_{4}$ & 9.72(0.14) & \textbf{9.18(0.14)} & 12.23(0.07) & 9.40(0.14) & \textbf{9.19(0.14)} & 12.16(0.07) & 9.58(0.13) & \textbf{9.18(0.14)}\\ 
					$ \hat{\psi}_{5}$ & 9.46(0.15) & \textbf{8.38(0.14)} & 12.19(0.07) & 8.97(0.15) & \textbf{8.50(0.14)} & 12.22(0.07) & 9.42(0.14) & \textbf{8.58(0.14)}\\ 
					\hline
					\# Conv & 413 & 500 & 45 & 500 & 500 & 37 & 484 & 500 \\
					\hline 
					\hline 
					 \multicolumn{9}{c}{$N =500$  } \\
					 \hline
					$ \varepsilon \sim \text{Normal}$  & REML & CG   & REML & EM & CG  & REML & EM & CG \\
					\hline
					$ \hat{\psi}_{1}$ & \textbf{1.81(0.04)} & 1.98(0.04) & 7.00(0.22) & 3.59(0.17) & \textbf{1.98(0.04)} & 6.11(0.21) & 3.79(0.17) & \textbf{1.98(0.04)}\\ 
					$ \hat{\psi}_{2}$ & \textbf{3.34(0.08)} & 3.54(0.08) & 8.38(0.20) & 4.69(0.15) & \textbf{3.54(0.08)} & 7.43(0.20) & 5.13(0.15) & \textbf{3.56(0.08)}\\ 
					$ \hat{\psi}_{3}$ & \textbf{5.25(0.11)} & 5.73(0.11) & 9.43(0.17) & 6.21(0.14) & \textbf{5.85(0.11)} & 8.49(0.17) & 6.94(0.14) & \textbf{5.81(0.11)}\\ 
					$ \hat{\psi}_{4}$ & \textbf{6.35(0.14)} & 7.01(0.15) & 9.94(0.15) & 7.42(0.16) & \textbf{7.09(0.15)} & 9.07(0.17) & 8.00(0.15) & \textbf{7.04(0.15)}\\ 
					$ \hat{\psi}_{5}$ & \textbf{5.12(0.14)} & 5.72(0.15) & 9.31(0.18) & 6.58(0.18) & \textbf{5.79(0.15)} & 8.20(0.19) & 7.01(0.17) & \textbf{5.77(0.15)}\\ 
					\hline
					\# Conv & 481 & 500 & 223 & 500 & 500 & 265 & 500 & 500\\
					\hline
					$\varepsilon \sim t_3$ & \\
					\hline
					$ \hat{\psi}_{1}$ & \textbf{1.77(0.03)} & 1.93(0.03) & 7.48(0.22) & 3.10(0.15) & \textbf{1.92(0.03)} & 6.02(0.21) & 3.33(0.14) & \textbf{1.93(0.03)}\\ 
					$ \hat{\psi}_{2}$ & \textbf{3.34(0.09)} & 3.56(0.08) & 8.64(0.20) & 4.29(0.14) & \textbf{3.57(0.08)} & 7.63(0.20) & 4.81(0.14) & \textbf{3.59(0.08)}\\ 
					$ \hat{\psi}_{3}$ & \textbf{5.54(0.13)} & 5.86(0.12) & 9.39(0.17) & 6.35(0.15) & \textbf{5.98(0.13)} & 8.76(0.17) & 6.90(0.14) & \textbf{5.94(0.13)}\\ 
					$ \hat{\psi}_{4}$ & \textbf{6.58(0.15)} & 6.97(0.15) & 9.79(0.16) & 7.19(0.16) & \textbf{7.04(0.15)} & 9.20(0.17) & 7.72(0.15) & \textbf{7.02(0.15)}\\ 
					$ \hat{\psi}_{5}$ & \textbf{5.21(0.14)} & 5.63(0.14) & 9.55(0.18) & 6.25(0.17) & \textbf{5.72(0.14)} & 8.39(0.19) & 6.78(0.17) & \textbf{5.70(0.14)}\\ 
					\hline
					\# Conv & 486 & 500 & 228 & 500 & 500 & 261 & 500 & 500 \\
					\hline
					$ \varepsilon \sim \text{Unif}$ & \\
					\hline
					$ \hat{\psi}_{1}$ & \textbf{1.76(0.03)} & 1.93(0.04) & 6.91(0.22) & 3.38(0.16) & \textbf{1.93(0.03)} & 6.05(0.22) & 3.55(0.16) & \textbf{1.93(0.04)}\\ 
					$ \hat{\psi}_{2}$ & \textbf{3.33(0.08)} & 3.57(0.08) & 8.29(0.19) & 4.55(0.15) & \textbf{3.57(0.08)} & 7.46(0.20) & 5.03(0.15) & \textbf{3.60(0.08)}\\ 
					$ \hat{\psi}_{3}$ & \textbf{5.46(0.12)} & 5.81(0.12) & 9.37(0.17) & 6.22(0.14) & \textbf{5.92(0.12)} & 8.80(0.18) & 6.76(0.14) & \textbf{5.90(0.12)}\\ 
					$ \hat{\psi}_{4}$ & \textbf{6.29(0.14)} & 6.68(0.14) & 9.57(0.16) & 6.95(0.16) & \textbf{6.78(0.14)} & 9.29(0.17) & 7.56(0.15) & \textbf{6.73(0.14)}\\ 
					$ \hat{\psi}_{5}$ & \textbf{4.91(0.13)} & 5.37(0.13) & 8.95(0.18) & 6.01(0.17) & \textbf{5.47(0.14)} & 8.36(0.19) & 6.64(0.16) & \textbf{5.43(0.14)}\\ 
					\hline
					\# Conv  & 478 & 500 & 219 & 500 & 500 & 266 & 500 & 500 \\
					\hline
					\hline
				\end{tabular}
			}
	\end{table}
	
	
	\subsection{Comparison between Knots Selection and Roughness Penalty}  \label{sec:numerics:penalty}

	\begin{table}[h!]
		\centering
		\caption{Estimation result  for the \textit{pracNU} setting. Our proposed method is tuned by selecting the number of knots (CG) or selecting the roughness tuning parameter (CG-p) with $15$ fixed number of knots. The reported values ($\times 10^{-1}$) are the average estimation errors for each eigenfunction $\|\hat{\psi}_r - \psi_{0r}\|_{L_2}$  based on 500 replications.  The reported numbers are the actual values multiplied by 10. The numbers in the parentheses are the standard errors. The first row indicates the initialization methods of EM, LS, and LOC. \label{tab:penaltyvsknotsNU}}
		\vspace{3pt}
		\resizebox{ \textwidth}{!}{
			\begin{tabular}{c|c c | c c | c c }
				\hline
				&  \multicolumn{2}{c}{EM  ($\times 10^{-1}$) } & \multicolumn{2}{|c|}{LS  ($\times 10^{-1}$)} & \multicolumn{2}{c}{LOC   ($\times 10^{-1}$) }\\
				\hline \hline
				\multicolumn{7}{c}{$N = 100$}   \\
				\hline  
				$ \varepsilon \sim \text{Normal}$   	&  CG  & CG-p& CG & CG-p& CG & CG-p  \\
				\hline
				$ \hat{\psi}_{1}$ & 5.91(0.14) & \textbf{5.76(0.14)} &   5.84(0.14) & \textbf{ 5.64(0.13)} &   5.85(0.13) & \textbf{ 5.66(0.13)}\\ 
				$ \hat{\psi}_{2}$ & 8.77(0.13) & \textbf{7.67(0.13)} &   8.74(0.13) & \textbf{ 7.83(0.14)} &   8.74(0.13) & \textbf{ 7.84(0.14)}\\ 
				$ \hat{\psi}_{3}$ & 9.66(0.13) & \textbf{8.77(0.14)} &   9.63(0.13) & \textbf{ 8.96(0.14)} &   9.65(0.13) & \textbf{ 8.97(0.14)}\\ 
				$ \hat{\psi}_{4}$ & 9.72(0.13) & \textbf{9.30(0.14)} &   9.62(0.13) & \textbf{ 9.28(0.13)} &   9.69(0.13) & \textbf{ 9.28(0.14)}\\ 
				$ \hat{\psi}_{5}$ & 8.86(0.14) & \textbf{8.40(0.14)} &   8.67(0.13) & \textbf{ 8.39(0.14)} &   8.74(0.13) & \textbf{ 8.47(0.14)}\\ 
				\hline
				$ \varepsilon \sim t_3$   &  \\
				\hline
				$ \hat{\psi}_{1}$ & 5.70(0.12) & \textbf{5.29(0.12)} &   5.60(0.12) & \textbf{ 5.37(0.12)} &   5.64(0.12) & \textbf{ 5.42(0.12)}\\ 
				$ \hat{\psi}_{2}$ & 8.91(0.13) & \textbf{7.52(0.13)} &   8.80(0.13) & \textbf{ 7.86(0.13)} &   8.76(0.13) & \textbf{ 7.90(0.14)}\\ 
				$ \hat{\psi}_{3}$ & 10.06(0.12) & \textbf{9.15(0.13)} & 10.01(0.12) & \textbf{ 9.19(0.13)} & 10.03(0.12) & \textbf{ 9.22(0.13)}\\ 
				$ \hat{\psi}_{4}$ & 9.84(0.13) & \textbf{9.44(0.13)} &   9.72(0.13) & \textbf{ 9.55(0.13)} &   9.80(0.13) & \textbf{ 9.44(0.14)}\\ 
				$ \hat{\psi}_{5}$ & 8.95(0.14) & \textbf{8.65(0.14)} &   8.83(0.14) & \textbf{ 8.74(0.15)} &   8.85(0.14) & \textbf{ 8.76(0.15)}\\ 
				\hline
				$ \varepsilon \sim \text{Unif}$   &  \\
				\hline
				$ \hat{\psi}_{1}$ & 5.66(0.13) & \textbf{5.44(0.13)} &   5.58(0.13) & \textbf{ 5.31(0.12)} &   5.56(0.12) & \textbf{ 5.32(0.12)}\\ 
				$ \hat{\psi}_{2}$ & 8.90(0.13) & \textbf{7.89(0.13)} &   8.87(0.13) & \textbf{ 7.84(0.13)} &   8.80(0.13) & \textbf{ 7.84(0.13)}\\ 
				$ \hat{\psi}_{3}$ & 10.02(0.13) & \textbf{9.05(0.14)} &   9.92(0.13) & \textbf{ 9.10(0.14)} &   9.77(0.13) & \textbf{ 9.23(0.14)}\\ 
				$ \hat{\psi}_{4}$ & 9.84(0.13) & \textbf{9.40(0.13)} &   9.83(0.13) & \textbf{ 9.37(0.14)} &   9.74(0.13) & \textbf{ 9.41(0.14)}\\ 
				$ \hat{\psi}_{5}$ & 9.19(0.14) & \textbf{8.88(0.15)} &   9.06(0.14) & \textbf{ 8.80(0.14)} &   8.98(0.14) & \textbf{ 8.76(0.15)}\\ 
				\hline
				\hline
				  \multicolumn{7}{c}{$N = 500$} \\
				\hline
				$\varepsilon \sim \text{Normal}$  	&  CG  & CG-p& CG & CG-p& CG & CG-p  \\
				\hline
				$ \hat{\psi}_{1}$ & 2.12(0.05) & \textbf{1.99(0.04)} &   2.10(0.05) & \textbf{ 1.99(0.04)} &   2.11(0.05) & \textbf{ 2.00(0.04)}\\ 
				$ \hat{\psi}_{2}$ & 3.87(0.09) & \textbf{3.55(0.08)} &   3.85(0.09) & \textbf{ 3.54(0.08)} &   3.83(0.09) & \textbf{ 3.53(0.08)}\\ 
				$ \hat{\psi}_{3}$ & 5.34(0.12) & \textbf{5.19(0.12)} &   5.39(0.12) & \textbf{ 5.28(0.12)} &   5.36(0.12) & \textbf{ 5.24(0.12)}\\ 
				$ \hat{\psi}_{4}$ & \textbf{6.16(0.13)} & 6.30(0.14) & \textbf{ 6.20(0.14)} &   6.40(0.14) & \textbf{ 6.19(0.14)} &   6.36(0.14)\\ 
				$ \hat{\psi}_{5}$ & \textbf{4.82(0.13)} & 4.99(0.13) & \textbf{ 4.83(0.13)} &   5.02(0.13) & \textbf{ 4.81(0.13)} &   5.00(0.13)\\ 
				\hline
				$\varepsilon \sim t_3$   &  \\
				\hline
				$ \hat{\psi}_{1}$ & 1.99(0.04) & \textbf{1.95(0.04)} &   1.98(0.04) & \textbf{ 1.94(0.04)} &   1.99(0.04) & \textbf{ 1.94(0.04)}\\ 
				$ \hat{\psi}_{2}$ & 3.77(0.08) & \textbf{3.59(0.08)} &   3.76(0.08) & \textbf{ 3.58(0.08)} &   3.75(0.08) & \textbf{ 3.59(0.08)}\\ 
				$ \hat{\psi}_{3}$ & 5.46(0.12) & \textbf{5.46(0.12)} & \textbf{ 5.53(0.12)} &   5.56(0.12) & \textbf{ 5.50(0.12)} &   5.53(0.12)\\ 
				$ \hat{\psi}_{4}$ & \textbf{6.29(0.14)} & 6.35(0.14) & \textbf{ 6.30(0.14)} &   6.43(0.14) & \textbf{ 6.27(0.14)} &   6.40(0.15)\\ 
				$ \hat{\psi}_{5}$ & \textbf{4.96(0.13)} & 5.11(0.13) & \textbf{ 4.96(0.13)} &   5.13(0.14) & \textbf{ 4.95(0.13)} &   5.12(0.13)\\ 
				\hline
				$ \varepsilon \sim \text{Unif}$   &  \\
				\hline
				$ \hat{\psi}_{1}$ & 1.98(0.03) & \textbf{1.92(0.04)} &   1.97(0.03) & \textbf{ 1.91(0.04)} &   1.98(0.03) & \textbf{ 1.92(0.04)}\\ 
				$ \hat{\psi}_{2}$ & 3.78(0.08) & \textbf{3.52(0.07)} &   3.80(0.08) & \textbf{ 3.51(0.07)} &   3.79(0.08) & \textbf{ 3.50(0.07)}\\ 
				$ \hat{\psi}_{3}$ & 5.20(0.12) & \textbf{5.13(0.12)} &   5.29(0.12) & \textbf{ 5.21(0.12)} &   5.27(0.12) & \textbf{ 5.16(0.12)}\\ 
				$ \hat{\psi}_{4}$ & \textbf{6.13(0.14)} & 6.20(0.14) & \textbf{ 6.18(0.14)} &   6.26(0.14) & \textbf{ 6.16(0.14)} &   6.23(0.14)\\ 
				$ \hat{\psi}_{5}$ & \textbf{4.94(0.13)} & 5.03(0.13) & \textbf{ 4.96(0.13)} &   5.06(0.13) & \textbf{ 4.95(0.13)} &   5.05(0.14)\\ 
				\hline
				\hline
			\end{tabular}
		}
	\end{table}

All the above methods are tuned by selecting the number of knots $K$ in the domain. 
In this subsection, we consider an alternative approach of penalized spline \citep{o1986statistical, o1988fast, huang2021asymptotic}, which fixes the number of knots at a relatively large value and a roughness penalization is introduced to trade off bias and variance. 
A large value of $K$ allows us to approximate more complex functions; meanwhile, the variance is controlled by the roughness penalty. 
	
The penalty measures the roughness of a function through the integral of its squared $d$-th $(1\le d \le O+1)$ derivative, i.e.,
\begin{equation} \label{eqn:penalty}
	\mathcal{P}_\eta(\mU) := 
	\eta \sum_{r=1}^{R}  \int \big\{\psi^{(d)}_{r}(u)\big\}^2
	\intd u = \eta  \tr\big( \mU\trans \mGamma \mU\big),
\end{equation}
where $\mGamma = \int \vb^{(d)}(u) \{\vb^{(d)}(u)\}\trans \intd u$ is a $K\times K$ matrix solely depending on the basis functions and $\eta$ is the penalty parameter. 
Note this penalty on the eigenfunctions can be transferred to the optimization parameter $\mU$ and thus can be incorporated into our method directly.
More precisely, 
combining the loss~\eqref{eqn:loss:logdet} and the penalty~\eqref{eqn:penalty}, the objective function becomes $\ell_\eta(\mU,\mW,\sigma_e^2) :=\mathcal{L}(\mU,\mW,\sigma_e^2) +\mathcal{P}_\eta(\mU,\mW)$.
We get our estimator from the $N$ sparse samples via solving
\begin{equation*}
	\begin{aligned}
		\min_{\mU, \mW,\sigma_e^2}\, & \ell_{\eta}(\mU,\mW,\sigma_e^2), \\
		\text{subject to }\, & \mU\in \mathrm{St}(R, K),\;
		\mW = \diag(\lambda_{1},\cdots, \lambda_{R})\succ \vzero, \text{ and } \sigma_e^2 >0 .
	\end{aligned}
\end{equation*}
The above optimization problem can still be solved by the proposed conjugate gradient algorithm over the product manifold, and we denote the corresponding estimator as CG-p in the following.
In this numerical experiment, we fix the number of knots as $15$ and select optimal $\eta$ by CV. 
	
The two approaches (CG and CG-p) are compared over a simulation setting named \textit{pracNU}, where a non-linear transformation is applied to the eigenfunctions $\psi_r$ in the $\textit{prac}$ setting of \cite{peng2009geometric}. 
Specifically, our experiment uses the eigenfunction $\Tilde{\psi}_{r}(u) = \psi_{r}(F(u; \alpha, \beta))$ for $r = 1, \ldots, 5$, where $F(u; \alpha, \beta)$ is the cumulative distribution function of beta distribution~(the regularized incomplete beta function) and the parameters are specified as $\alpha = \beta = 2$. 
After the non-linear transformation, the shape of eigenfunction becomes more complicated, and it's interesting to compare the performance of discrete tuning (the number of knots) and continuous tuning (roughness penalization). The results of this simulation study under various error distributions and sample sizes are summarized in Table \ref{tab:penaltyvsknotsNU}.
	
Table~\ref{tab:penaltyvsknotsNU} shows some advantages of CG-p over CG. 
When the sample size is $100$, the estimation errors of functional principal components have been reduced when the roughness penalization is used. 
However, when the sample size is relatively large ($N=500$), the two methods have competitive performance. We thus recommend using CG-p when the sample size is relatively small.
Note that CG-p could possibly be improved further by assigning different tuning parameters for each principal component function.

\subsection{Real Data Analysis}  \label{sec:numerics:realdata}	
	
\begin{figure}[h!]
	\centering
	\includegraphics[width = 0.55\textwidth]{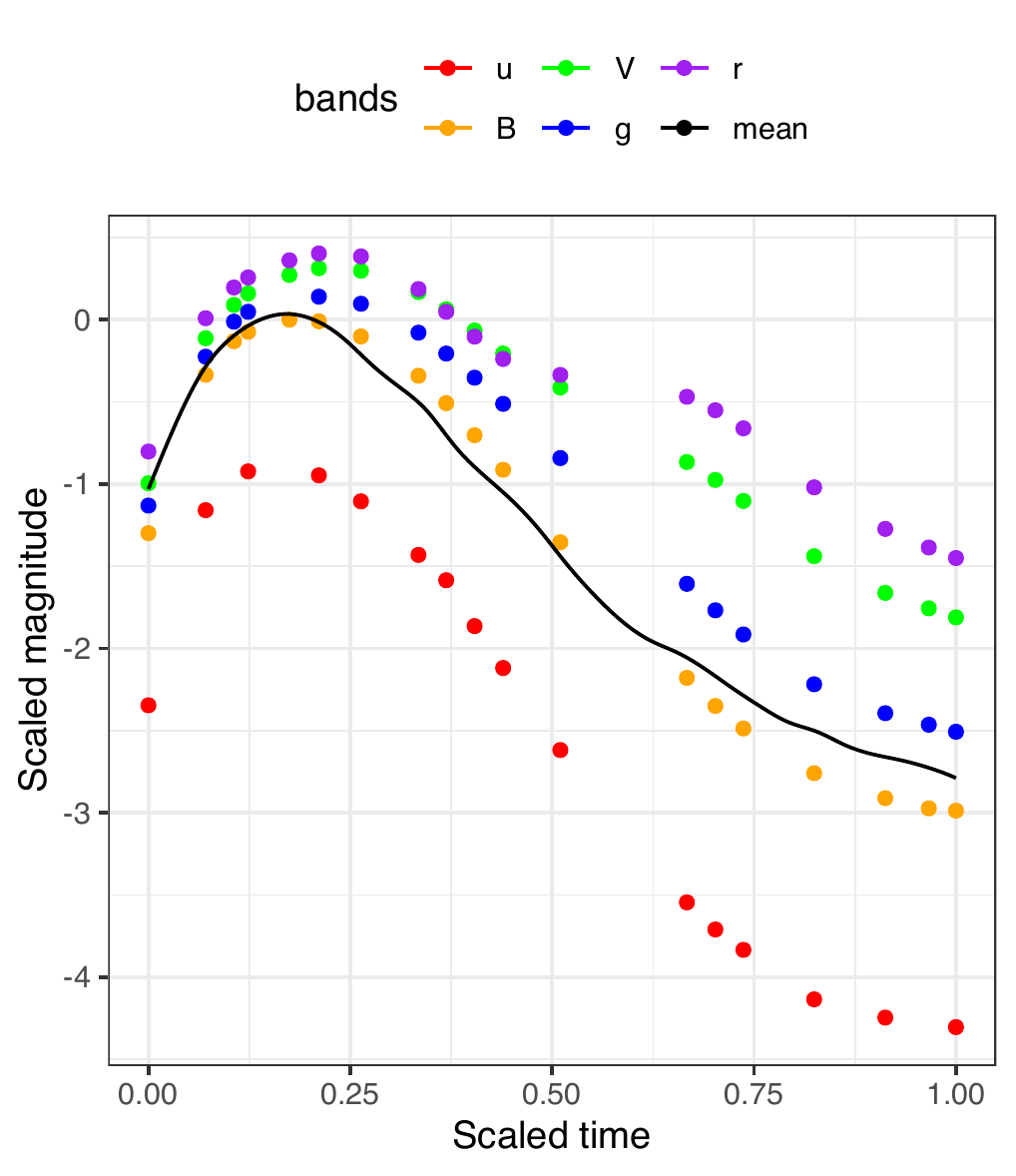}
	\caption{An example of Type Ia supernova light curve data. The points are the observed data from one supernova from distinct bands and the solid curve is the estimated mean curve. 
	\label{fig:mean_curve}}
\end{figure}
	
We further compare these methods on the Type Ia supernova light curve data obtained from the Carnegie Supernova Project \citep[CSP,][]{stritzinger2011carnegie}. Type Ia supernova (SNe) is an important ``standardizing'' candle for us to probe the universe and measure the distance. 
In the CSP dataset, each supernova is observed by nine bands (or filters), which only allow the light within a certain wavelength range to pass through. Different bands are usually denoted by distinct letters ($u,g,r,B,V$, etc.). 
Astronomers typically use the peak magnitude (brightness) to measure distances. 
	
We only selected the SNe samples which had measurement before the peak $B$-band magnitude. This allowed us to perform curve registration and align the peak magnitude to the origin point of the plane.
{
Specifically,  a smooth spline function was fitted to the $B$-band observation of each sample, and the peak magnitude time was identified from the fitted function. The observation time was then subtracted by the peak magnitude time such that the peak magnitude got registered to time $0$.}
After the curve registration, we only kept the observation points within $-10$ to $+40$ days around the peak.  The range of observational time points was then scaled to the $[0,1]$. 
We treated the light curves from different bands as independent samples. The total number of resulting sample curves is $N = 258$. 
The registered and scaled light curves of one supernova are depicted in Figure~\ref{fig:mean_curve}, where various colored points represent sparse observations from distinct bands. 
Before using FPCA to fit the data, a common mean curve was firstly estimated and subtracted from the data. The fitted mean curve, denoted by $\hat{\mu}$, is also shown in Figure~\ref{fig:mean_curve} as a solid curve.

\begin{figure}[!h]
	\centering
	\includegraphics[width = 0.9\textwidth]{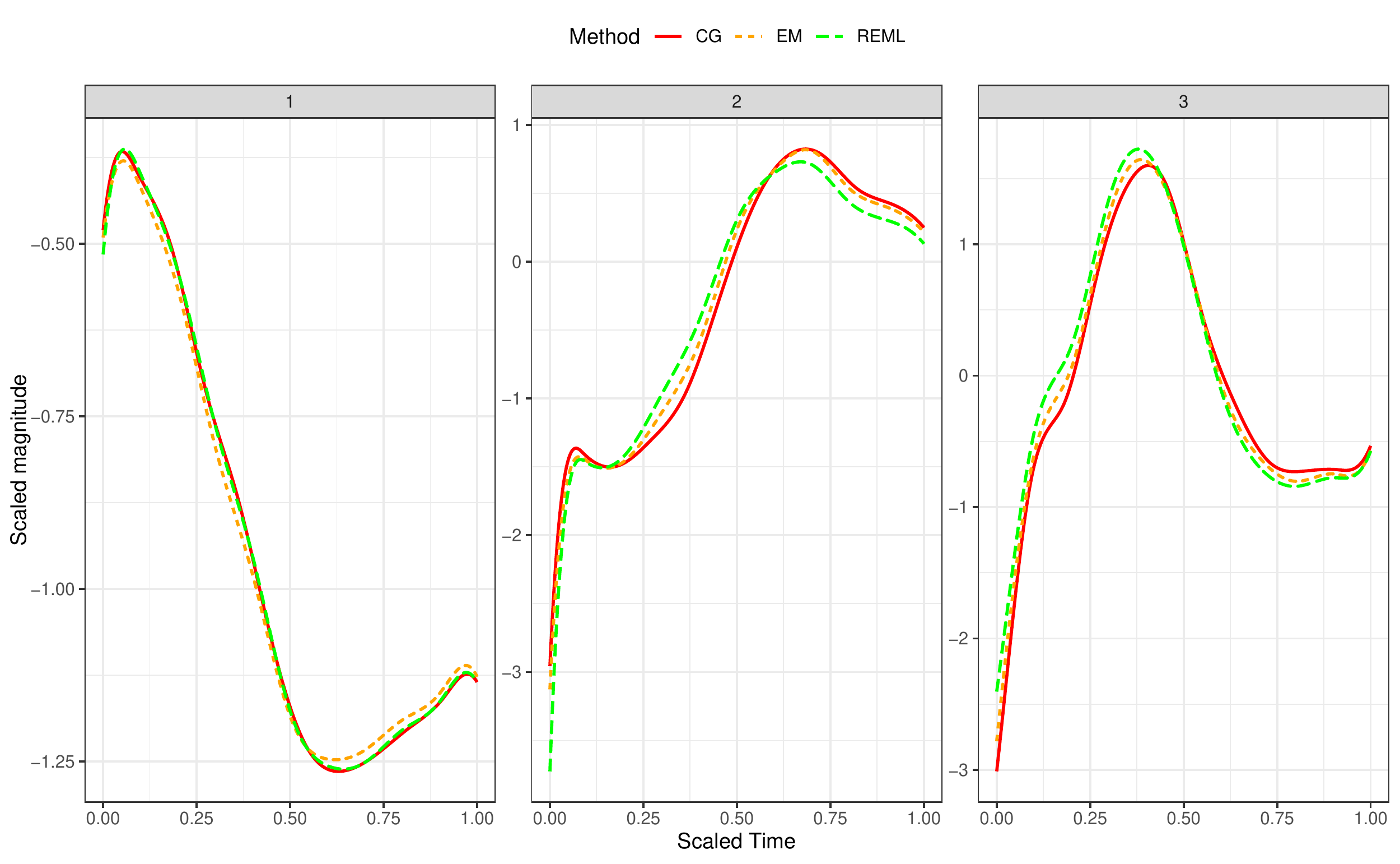}
	\caption{The three panels from left to right correspond  to  the  leading  three estimated eigenfunctions by CG, EM, and REML.}
	\label{fig:pc}
\end{figure}

The proposed CG and the alternative methods discussed in the previous section, i.e., EM and REML, were applied to the pre-processed data to estimate the first three principal component functions. 
The methods were all tuned by CV to select the number of knots. 
The estimation results of the first 3 eigenfunctions by CG, EM, and REML are depicted in Figure~\ref{fig:pc}. It shows that the estimated eigenfunctions by all methods roughly follow the same pattern.

Since the true eigenfunctions are unknown in the real data, we accessed the prediction accuracy to compare different methods. In this part, we also included CG-p, the conjugate gradient method with roughness penalty, as a competitor.
The dataset got randomly split into two subsets of curves, i.e., a training set and a test set. They account for 80\% and 20\% of the total curves, respectively.
On the training dataset, $R(=2,3,4)$ principal components were estimated by each FPCA method. Let $\hat{\psi}_r$'s and $\hat{\lambda}_r$'s be respectively the estimated eigenfunctions and eigenvalues, $r=1,\dots,4$, and $\hat{\sigma}_{e}^{2}$ be the estimated noise variance. 
	
The prediction accuracy was evaluated on the test dataset as follows. For each light curve in the test set, we randomly selected 50\% of its observational points to predict its scores and fitted the light curve. The remaining points were set as hold-out data. In other words, for the $n$-th sample with $M_n$ points, the index set $\{1,\cdots, M_n\} = O_n \cup H_n$ was randomly split into two disjoint subsets $O_n$ and $H_n$ of roughly the same size. The two sets correspond to the observed part and the held-out part, respectively.
Suppose we have observed $\vy_{n*} = (y_{nj})_{j\in O_n}\trans$ at the time points $\{u_{nj}\}_{j\in O_n}$. The corresponding mean vector is denoted as $\hat{\boldsymbol{\mu}}_{n*} = [\hat{\mu} (u_{nj})]_{j\in O_n}\trans$, with $\hat{\mu}$ being the estimated mean function $\mu$ in the data pre-processing step. 
Define ${\mPsi}_{n*} = [{\psi}_r(u_{nj})]_{j\in O_n, r}$ as a matrix of size $\vert O_n\vert \times R$. The $r$-th column of ${\mPsi}_{n*}$ contains the evaluation of the $r$-th eigenfunction at the observed time points, $r=1,\dots, R$. 
	
\begin{figure}[h!]
	\centering
		\includegraphics[width=0.45\textwidth]
		{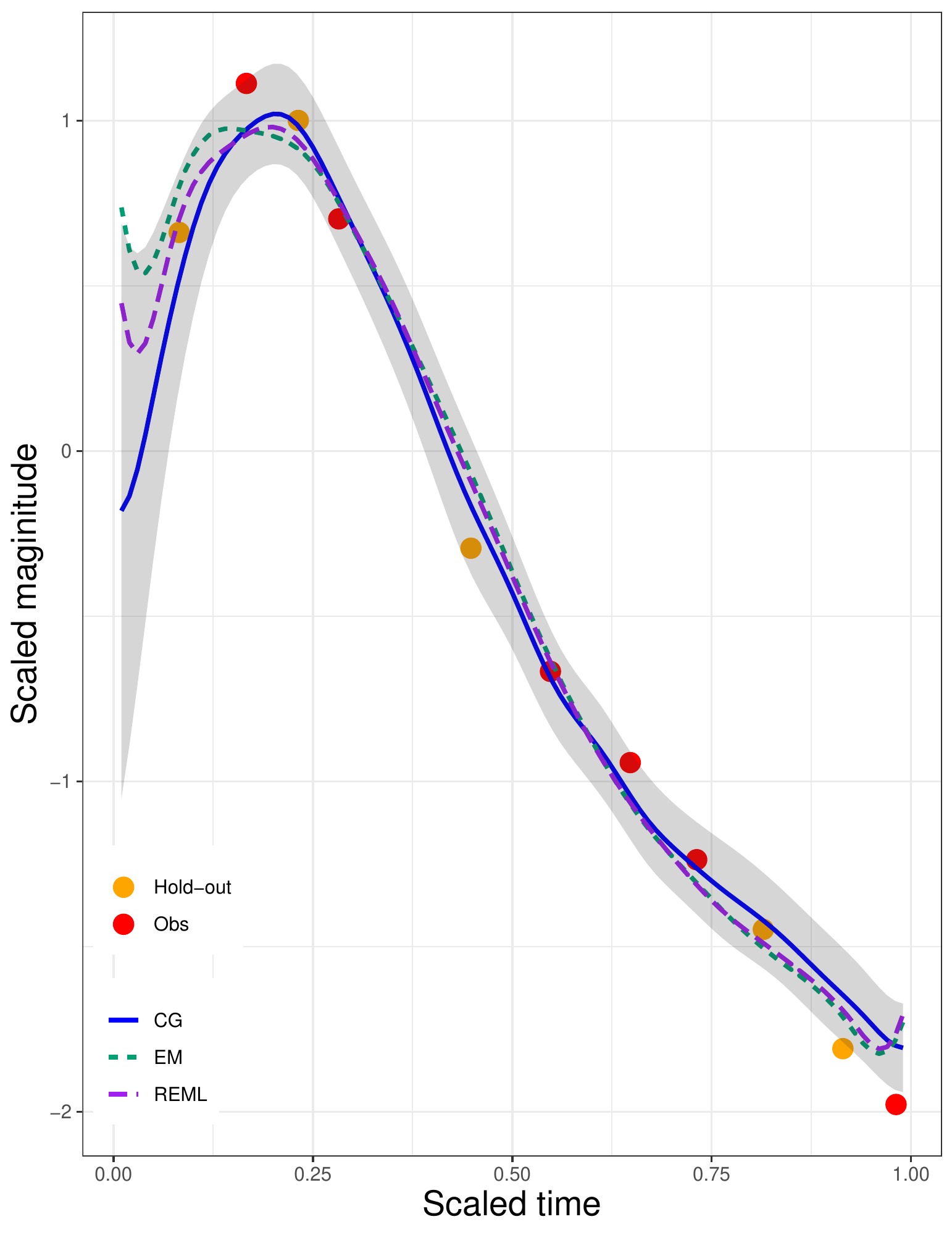}
	\hfill
		\includegraphics[width=0.45\textwidth]
		{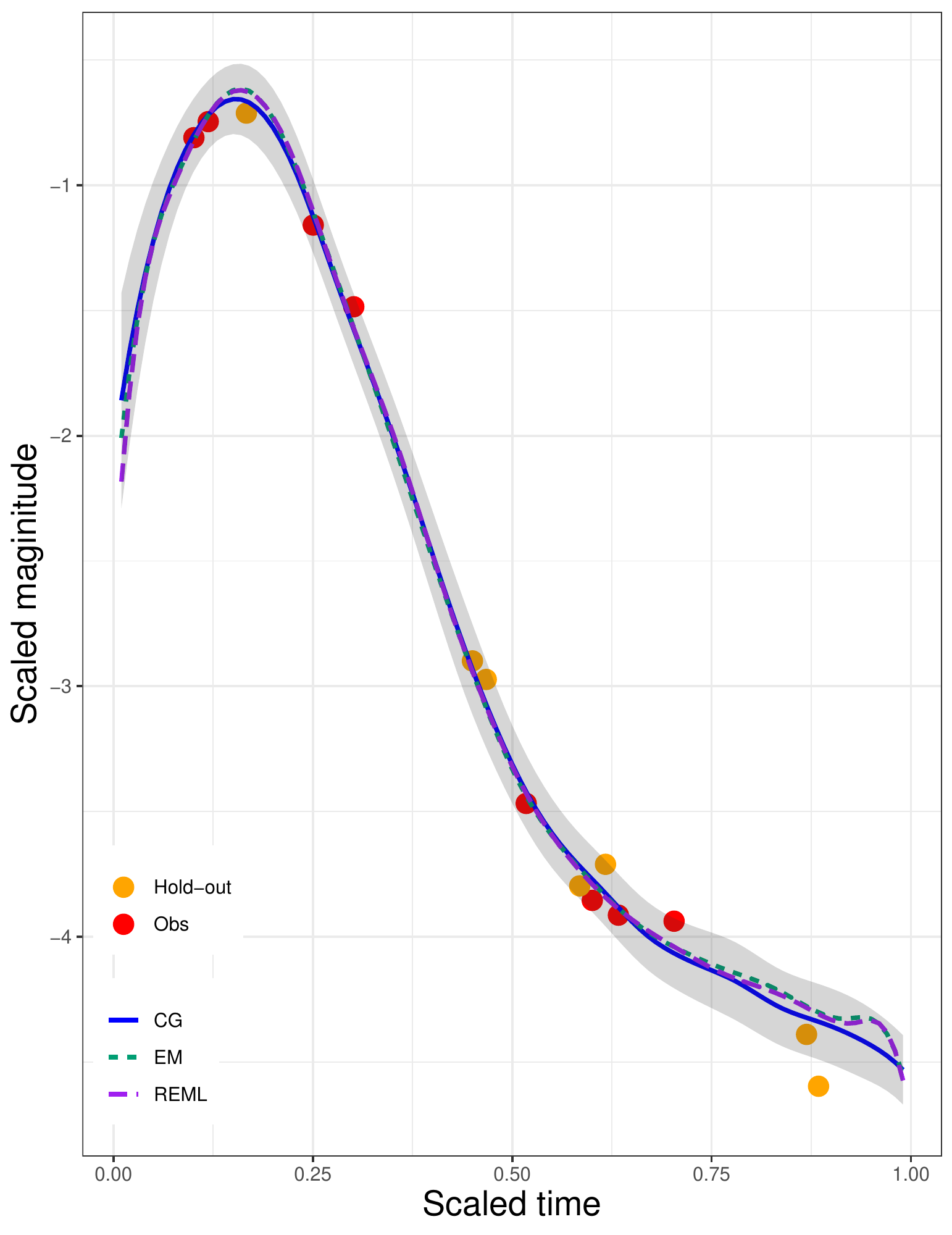}
	\caption{Two examples of curve fitting by the three methods.
	The red dots are the observed points in the test curves that we used to predict the PC scores. The yellow dots are held-out data points for which we test the prediction performance. The predicted light curves by the proposed CG method are shown as blue solid lines in the center. The corresponding 95\% predictive intervals by CG are shown as grey areas. The predicted curves by EM and REML are shown in green short dashed and purple long dashed, respectively.
	\label{fig:pred}}
\end{figure}
	
\begin{table}[h!]
	\caption{Prediction performance on the real data. Reported are the average MSPE ($\times 10^{-2}$)  over the test datasets for REML, EM, CG, and CG-p. The reported numbers are the actual values multiplied by 100.
	The first row indicates the initialization methods of EM, LS, and LOC. The blocks correspond to different model ranks $R$.
	The numbers in the parentheses are the standard errors. 
	Totally $500$ replicates are conducted.\label{tab:MSFE}}
	\vspace{3pt}
	\centering
	\resizebox{ \textwidth}{!}{
		\begin{tabular}{ c c c  |c c c c| c c c c }
			\hline \noalign{\smallskip}
			\multicolumn{3}{c|}{EM ($\times 10^{-2}$ ) } & \multicolumn{4}{|c|}{LS ($\times 10^{-2}$ ) } & \multicolumn{4}{c}{LOC ($\times 10^{-2}$ ) } \\
			\hline \noalign{\smallskip}
			REML & CG & CG-p& REML &EM   & CG & CG-p & REML &EM   & CG & CG-p\\
			\hline \noalign{\smallskip}
			\hline	\multicolumn{11}{c}{$R=2$ }	 \\ \hline
			2.790 & 2.764 & \textbf{2.744} & 2.962 & 2.783 & 2.748 & \textbf{2.715} & 3.076 & 2.782 & 2.759 & \textbf{2.737}\\ 
			(0.029) & (0.027) & (0.027) & (0.155) & (0.028) & (0.027) & (0.026) & (0.037) & (0.028) & (0.027) & (0.027)\\ 
			\hline \hline	\multicolumn{11}{c}{$R=3$ }	 \\ \hline
			1.756 & 1.696 & \textbf{1.649} & 7.665 & 1.710 & 1.684 & \textbf{1.642} & 3.087 & 1.844 & 1.799 & \textbf{1.657}\\ 
			(0.024) & (0.022) & (0.021) & (0.649) & (0.022) & (0.022) & (0.021) & (0.097) & (0.033) & (0.032) & (0.023)\\ 
			\hline \hline	\multicolumn{11}{c}{$R=4$ }	 \\ \hline
			1.285 & 1.278 & \textbf{1.269} & 11.153 & 1.290 & 1.269 & \textbf{1.258} & 9.599 & 1.299 & 1.272 & \textbf{1.257}\\ 
			(0.018) & (0.018) & (0.018) & (0.830) & (0.019) & (0.018) & (0.018) & (0.841) & (0.019) & (0.018) & (0.018)\\  
			\hline
			\hline
		\end{tabular}
	}
\end{table}
	
Let $\theta_{nr}$ be the PC scores of $n$-th sample in the $r$-th component. Denote $\bm{\theta_{n}} = (\theta_{n1},\dots, \theta_{nR})\trans$. We can see the joint vector $(\vy_{n*}\trans,\bm{\theta}_{n}\trans)\trans $ of the observations and the scores has the following first and second moments
\begin{equation}\label{equ:cond_normal}
	\Expect  \begin{pmatrix}
		\vy_{n*} \\
		\bm{\theta}_{n}
	\end{pmatrix} =
	\begin{pmatrix}
		{\boldsymbol{\mu}}_{n*} \\
		\bm{0}
	\end{pmatrix},\quad
	\Var  \begin{pmatrix}
		\vy_{n*} \\
		\bm{\theta}_{n}
	\end{pmatrix} =
	\begin{pmatrix}
		{\mPsi}_{n*}{\mW}{\mPsi}_{n*}\trans + {\sigma}_{e}^{2}\mI & \ \ {\mPsi}_{n*}{\mW} \\
		{\mW}{\mPsi}_{n*}\trans & {\mW}
	\end{pmatrix}.
\end{equation}
Plug-in the estimated $\hat{\boldsymbol{\mu}}_{n*}$,  $\hat{\sigma}_{e}^{2}$, $\hat{\mW}$, and  $\hat{\mPsi}_{n*}$, \eqref{equ:cond_normal} leads us to the best linear unbiased prediction (BLUP) of the score $\bm{\theta}_{n}$ as 
$$
	\hat{\bm{\theta}}_{n} =  \hat{\mW}\hat{\mPsi}_{n*}\trans( \hat{\mPsi}_{n*} \hat{\mW}\hat{\mPsi}_{n*}\trans + \hat{\sigma}_{e}^{2}\mI)^{-1}(\bm{y}_{n*} - \hat{\boldsymbol{\mu}}_{n*} ).
$$
Given the predicted score $\hat{\bm{\theta}}_{n}$, the mean squared prediction errors~(MSPE) over the held-out observations indexed by $H_n$ is evaluated as
\begin{equation*}
	\text{MSPE} = 
	\frac{1}{\vert H_n\vert}\sum_{j\in H_n}
	\Big\{y_{nj} - \hat{\mu}(u_{nj}) - \sum_{r = 1}^{R} \hat{\lambda}_{r}^{1/2}\hat{\theta}_{nr}\hat{\psi}_{r}(u_{nj})\Big\}^{2}.
\end{equation*}
Figure~\ref{fig:pred} illustrates this predicting procedure for two curves in the test dataset. It plots the fitted curve by using $R=4$ FPCs estimated by  CG over the training dataset in one random splitting. 
The red and orange points correspond to the observed (indexed by $O_n$) and held-out (indexed by $H_n$) data points, respectively.
The solid blue lines are the fitted curves based on the red (observed) points, and then MSPE is evaluated for the held-out data points in orange.  The gray area represents the 95\% point-wise predictive interval. The predicted curves by EM (green short dashed) and REML (purple long dashed) are included for comparison.

We carried out $500$ random splitting of the above procedure. In each replication, the average MSPE using each method under consideration (among EM, REML, CG, and CG-p) was computed for all the held-out samples in the test dataset.
We tested various numbers $R (=2,3,4)$ of fitted functional principal components.
Table~\ref{tab:MSFE} reports the average MSPE under these random splittings with the corresponding standard errors. 
The results in Table~\ref{tab:MSFE} are consistent with those in the simulation studies. 
In particular, Table~\ref{tab:MSFE} shows that REML leads to a large prediction error when the initial value is not good enough. 
Our methods (CG and CG-p) enjoy the best prediction accuracy compared with the other methods in consideration. CG-p with a penalty also shows some improvement over CG.

{
\section{Discussion}\label{sec:discuss}

This paper has proposed a manifold conjugate gradient algorithm to estimate the functional principal components when the sample curves are sparse and irregularly observed.  Using basis expansion, we have formulated the objective function on the product space of the Stiefel manifold and the cone of positive definite matrices. 
The proposed method improves the search efficiency and estimation accuracy by taking into account the intrinsic geometric structure of the underlying product manifold. 
Our objective function is directly derived from a matrix Bregman divergence, and thus no distribution assumption is needed to apply the proposed method.
We also have shown that a roughness penalization can be easily incorporated into our algorithm with a better fit. 
The superior performance and numerical stability of the proposed method  against the existing ones are demonstrated by simulation studies and a real application on Type Ia supernova light curve data.

Our manifold conjugate gradient algorithm can be extended to general loss functions by choosing different seed functions in the matrix Bregman divergence. A comprehensive study on the comparison of different loss functions for functional principal component analysis is a potential future research topic. Furthermore, investigating the asymptotic properties of these estimators is also of interest.

\clearpage

\bibliography{reference}

%
%


	
	
	
\end{document}